\documentclass[%
 reprint,
 amsmath,amssymb,
pra,
]{revtex4-1}
\usepackage{latexsym}
\usepackage{amsmath,amssymb}
\usepackage{amsthm}
\usepackage{mathdots}
\makeatother%
\usepackage{braket}
\newcommand{\1}{\mbox{1}\hspace{-0.23em}\mbox{l}}
\usepackage{mathdots}
\usepackage{graphicx,color}% Include figure files
\usepackage{dcolumn}% Align table columns on decimal point
\usepackage{bm}
\usepackage{graphicx}
\usepackage{ulem}
\begin{document}
\preprint{APS/123-QED}

\title{$\mathcal{PT}$ phase transition in open quantum systems with Lindblad dynamics}% Force line breaks with \\

\author{Yuma Nakanishi}
 \altaffiliation%[Also at ]
 {}
  \email{nakanishi.y@stat.phys.titech.ac.jp}%Lines break automatically or can be forced with \\
\author{Tomohiro Sasamoto}%
\affiliation{%
Department of Physics, Tokyo Institute of Technology, 2-12-1 Ookayama,
Meguro-ku, Tokyo, 152-8551, JAPAN}%

\date{\today}% It is always \today, today,

\begin{abstract}
%Parity and time-reversal (PT) phase transitions in open systems with an exact balance between gain and loss have been widely studied in terms of non-Hermitian Hamiltonian. Here 
%Recent studies on PT symmetry in open quantum systems with Lindblad dynamics have been actively investigating although the definition of Liouvillian PT symmetry has not been decided yet.
We investigate parity-time ($\mathcal{PT}$) phase transitions in open quantum systems and discuss a criterion of Liouvillian $\mathcal{PT}$ symmetry proposed recently by Huber \textit{et al}. [J. Huber \textit{et al}., SciPost Phys. $\textbf{9}$, 52 (2020)]. Using the third quantization, which is a general method to solve the Lindblad equation for open quadratic systems, we show, with a proposed criterion of $\mathcal{PT}$ symmetry,  that the eigenvalue structure of the Liouvillian clearly changes at the $\mathcal{PT}$ symmetry breaking point for an open 2-spin model with exactly balanced gain and loss if the total spin is large. In particular, in a $\mathcal{PT}$ unbroken phase, some eigenvalues are pure imaginary numbers while in a $\mathcal{PT}$ broken phase, all the eigenvalues are real. From this result, it is analytically shown for an open quantum system including quantum jumps that the dynamics in the long time limit changes from an oscillatory to an overdamped behavior at the proposed $\mathcal{PT}$ symmetry breaking point. Furthermore, we show a direct relation between the criterion of Huber \textit{et al}. of Liouvillian $\mathcal{PT}$ symmetry and the dynamics of the physical quantities for quadratic bosonic systems. Our results support the validity of the proposed criterion of Liouvillian $\mathcal{PT}$ symmetry.% for the case where gain and loss (dissipations) are exactly balanced as in the PT symmetric Hamiltonian case.
\end{abstract}

\maketitle

%\tableofcontents

\section{\label{sec:level1}Introduction}
A variety of open classical systems, including mechanical systems $\text{\cite{Huberr}}$, optical systems $\text{\cite{Ruschhaupt}}$, electrical systems $\text{\cite{Joglekar}}$, can be described by non-Hermitian Hamiltonians (NHHs) $\text{\cite{Ashidasan}}$.
%An open system in the classical regime (i.e. without quantum effects) that interacts with the environment can be effectively described by non-Hermitian Hamiltonian (NHH) $\text{\cite{Ashidasan}}$.
Bender and Boettcher showed that, for a broad class of NHHs with parity-time ($\mathcal{PT}$) symmetry, phase transitions occur, in which eigenvalues change from real to complex $\text{\cite{BenderC.M.Boettcher}}$. This type of transitions are called $\mathcal{PT}$ phase transitions. It is also known that a $\mathcal{PT}$ symmetry breaking of eigenstates and a change of the nature of dynamics from oscillation to divergence or decay occur at the $\mathcal{PT}$ phase transition point. These abrupt changes have been observed in a variety of physical experimental system in mechanics $\text{\cite{Bender}}$, photonics $\text{\cite{RoterC}}$, %optomechanics $\text{\cite{Xu}}$, 
plasmonics $\text{\cite{Alaeian}}$, electronics $\text{\cite{Schindler}}$, and so on. %Here, it is known that non-Hermitian effects correspond to the dissipations of the system's energy, particles, and so on, which are called gain or loss.
Furthermore, various unconventional phenomena such as power oscillation $\text{\cite{RoterC,Makris}}$, enhancement of sensing $\text{\cite{Chen}}$, loss-induced transparency $\text{\cite{Guo}}$, non-reciprocal propagation $\text{\cite{Ramezani}}$ have also been observed in the vicinity of $\mathcal{PT}$ phase transition points.%, for example, power oscillation, enhancement of sensing, loss-induced transparency $\cite{Guo}$, non-reciprocal propagation $\cite{Ramezani,Peng}$, PT-symmetric lasers $\cite{Feng,Hodaei,Longhi}$, unidirectional invisibility $\cite{Lin}$, and so on.\\

In contrast, an open quantum system can be described by a Liouvillian, in particular by a Lindblad equation if the evolution of states is Markovian and completely positive 
trace preserving $\text{\cite{Lindbladref,Breuer,ARivas}}$. The Lindblad equation includes the effect of quantum jumps, which cause instantaneous switchings between energy levels in quantum systems. In certain situations the effect of quantum jumps can be ignored by using postselection, 
and then the time evolution of the system is described by NHHs $\text{\cite{MingantiH,Minganti1}}$. 
$\mathcal{PT}$ phase transitions for such open quantum systems without the effects of quantum jumps have also been observed $\text{\cite{Wu1,Naghiloo}}$, which are similar to open classical systems. However, quantum jumps often play an essential role, so for general open quantum systems we must consider the Lindblad equation. 

The study of phase transitions in open quantum systems, in particular dissipative phase transitions, has been actively conducted recently 
$\text{\cite{Minganti2,Kessler,Hwang,Casteels1,Rota1,Lee1}}$. However, the relation between the $\mathcal{PT}$ symmetry breaking and dissipative phase transitions has not been well understood. For example, a transition between purely real and complex eigenvalues of the Liouvillian has not been found even for the case where gain and loss (dissipations) are exactly balanced at a macroscopic level as in the $\mathcal{PT}$ symmetric Hamiltonian case. %Here, in open classical systems, NHHs have PT symmetry if gain and loss are exactly balanced. 
Also, a time reversal equivalence of gain and loss is broken by quantum noises at a microscopic level $\text{\cite{Scheel}}$. 
Therefore, a clear and well defined notion of Liouvillian $\mathcal{PT}$ symmetry has not been established. 
However recently there have been a few proposals for its definition including the one by Huber \textit{et al}. $ \text{\cite{Huber2}}$
(see also $\text{\cite{Prosen1}}$ for another) and a few works to try to clarify its meaning and validity.

In $\text{\cite{Huber2,Huber1}}$, the authors investigated a certain model with criterion of Huber et al. of Liouvillian $\mathcal{PT}$ symmetry in which a transition like a $\mathcal{PT}$ symmetry breaking occurs. Specifically, for an open 2-spin model with $XX$ interaction and exactly balanced gain and loss, it has been shown that a symmetry parameter $\text{\cite{Kepesidis}}$, which provides a measure for the parity symmetry of the density operator, changes from zero to a finite value suddenly at a point, where the proposed $\mathcal{PT}$ symmetry breaks down. 
This would be an indication that the proposed criterion is reasonable and the $\mathcal{PT}$ symmetry breaking indeed occurs in the system, but 
many characteristic features of the $\mathcal{PT}$ symmetry breaking in NHHs have not been clearly demonstrated. 
For example, a clear transition of the Liouvillian eigenvalue structure has not  been observed, and a direct relation between the dynamics and the criterion 
of Huber \textit{et al}. of Liouvillian $\mathcal{PT}$ symmetry has not been well understood. 

In this paper, we investigate $\mathcal{PT}$ phase transitions in open quantum systems, in particular, change of eigenvalue structure of the Liouvillian and the time dependence of physical quantities at a point where the $\mathcal{PT}$ symmetry breaks down according to the criterion of Huber \textit{et al}. 
As a method to find the eigenvalue structure, we will use the third quantization $\text{\cite{Prosen3,Prosen4}}$, which can be used for open quadratic fermionic or bosonic systems with linear bath operators. 

First, we will apply the third quantization to the open 2-spin model studied in Refs.$\text{\cite{Huber1,Huber2}}$ when the total spin $S$ is large. For a large $S$, we can linearly transform the spin ladder operators to bosonic annihilation and creation operators by the Holstein-Primakoff (HP) approximation $\text{\cite{Holstein}}$. Then we can apply the third quantization and, as a result, we will show that the transition between purely real and complex eigenvalues occurs at the $\mathcal{PT}$ symmetry breaking point according to the criterion of Huber \textit{et al}. Also, from this result, we will analytically show that the dynamics in the long time limit changes from an oscillatory to an overdamped behavior at the same transition point. 
This supports the proposed criterion of $\mathcal{PT}$ symmetry, and once we accept it, our results means that 
we will observe for the first time a clear change of the eigenvalue structure and the dynamics in the long time limit at the $\mathcal{PT}$ symmetry breaking point for the Liouvillian as in the $\mathcal{PT}$ symmetric Hamiltonian case.

Furthermore, we will apply the third quantization to general quadratic bosonic models with linear bath operators and show that, if the system satisfies the criterion of Huber \textit{et al}. of Liouvillian $\mathcal{PT}$ symmetry, a matrix $\textbf{X}$, which constitutes the Liouvillian for quadratic bosonic models, multiplied by the imaginary unit $i$, commutes with the operator $PT$. 
In other words, we show that criterion of Huber \textit{et al}., which was introduced based on phenomenological considerations, 
can be rewritten in the same form as the conventional condition of  $\mathcal{PT}$ symmetry. 

Also, we will show that the time dependence of the one point and two point correlation functions for quadratic bosonic systems can be determined by the eigenvalues of the matrix $\textbf{X}$. From these results, we will show that the physical quantities oscillate in the long time limit if the conventional $\mathcal{PT}$ symmetry of the matrix $i\textbf{X}$ is unbroken, while the physical quantities diverge if it is broken. In other words, we will show that criterion of Huber \textit{et al}. of Liouvillian $\mathcal{PT}$ symmetry directly relates the dynamics of the physical quantities for quadratic bosonic systems.

This paper is organized as follows: in section I$\hspace{-.1em}$I, we explain the conventional condition of $\mathcal{PT}$ symmetry, and the Liouvillian spectral properties and $\mathcal{PT}$ symmetry. Also, we review the bosonic third quantization. In section I$\hspace{-.1em}$I$\hspace{-.1em}$I, we analyze the eigenvalue structure and the dynamics for the open 2-spin model with the criterion of Huber \textit{et al}. of Liouvillian $\mathcal{PT}$ symmetry. Furthermore, we investigate the quantum fluctuation for the $\mathcal{PT}$ and $\mathcal{PT}$ broken phase by the quantum trajectory analysis. In section IV, we investigate the condition of the matrix $\textbf{X}$ from the criterion of Huber \textit{et al}. of Liouvillian $\mathcal{PT}$ symmetry and give the time dependence of the one point and two point correlation functions for the quadratic bosonic systems. In section V, we summarize this paper and state the outlook for further work. In Appendix A, we briefly explain the third quantization for the case where $\textbf{X}$ is not diagonalizable. In Appendix B, we explain the Holstein-Primakoff transformation briefly. In Appendix C, we give details of our calculations of the third quantization for the open 2-spin model. In Appendix D, we show the condition of the matrices, which constitute the matrix $\textbf{X}$, from the criterion of Huber \textit{et al}. of Liouvillian $\mathcal{PT}$ symmetry. In Appendix E, we derive the time derivatives of the one point and two point correlation functions for the quadratic bosonic systems. 
%\ts{[Add description of appendices.]}

\section{PT symmetry and Liouvillian spectrum}

\subsection{PT symmetric matrices and operators}
A matrix or a linear operator $A$ is said to be $\mathcal{PT}$ symmetric $\text{\cite{BenderC.M.Boettcher,MostafazadehA1}}$ if it commutes with the $PT$ operator that combines the parity operator $P$ and the time reversal operator $T$, that is,
\begin{eqnarray}
\label{NHHPT}
[A, PT]=0.
\end{eqnarray}
In this paper, this condition ($\ref{NHHPT}$) is called the conventional $\mathcal{PT}$ symmetry to distinguish it from Liouvillian $\mathcal{PT}$ symmetry.
The $\mathcal{PT}$ symmetry is said to be unbroken if all the eigenvectors of the $\mathcal{PT}$ symmetric $A$ are eigenstates of the $PT$ operator. In this case, all the eigenvalues of the $\mathcal{PT}$ symmetric $A$ are real. On the other hand, the PT symmetry is said to be broken if some eigenvectors of the $\mathcal{PT}$ symmetric $A$ are not eigenvectors of the 
$PT$ operator. In this case
some eigenvalues are complex conjugate pairs $\text{\cite{MostafazadehA1}}$. 

If $A$ appears in the Schr$\ddot{\textrm{o}}$dinger-type equation, 
\begin{eqnarray}
\label{Scheq}
i\frac{d}{dt}\psi=A\psi,
\end{eqnarray}
we call $A$ to be a Hamiltonian and often write it as $H$. When $H$ satisfies ($\ref{NHHPT}$) $\text{\cite{BenderC.M.Boettcher}}$, it is called a $\mathcal{PT}$ symmetric Hamiltonian, including the case where $H$ is not Hermitian.
From the above explanation, the solution $\psi$ of (\ref{Scheq}) is an oscillating solution if $\mathcal{PT}$ symmetry is unbroken, while there are some solutions that diverge or decay exponentially if $\mathcal{PT}$ symmetry is broken.

\subsection{Lindblad equation}
We consider open quantum systems
where the evolution of states is completely positive and trace preserving and Markovian. Then time evolution the density matrix $\rho(t)$ for the systems 
is described by Lindblad master equation $\text{\cite{Lindbladref,Breuer,ARivas}}$,
\begin{align}
\label{lindblad}
\frac{d\rho}{dt}=-i[H,\rho(t)]+\sum_{i}\Gamma_{i}\mathcal{D}[c_{i}]\rho,
\end{align}
where $H$ is a Hermitian Hamiltonian, $i$ represents the index of a bath, and the dissipation superoperators
$\mathcal{D}[c_{i}]$ are defined as
\begin{align}
\mathcal{D}[c_{i}]\rho=2c_{i}\rho c_{i}^{\dagger}-c_{i}^{\dagger}c_{i}\rho-\rho c_{i}^{\dagger}c_{i}.
\label{D}
\end{align}
Here, $c_{i}$ and $\Gamma_{i}$ are the Lindblad operator and dissipation rate respectively, and $\Gamma_{i}$ is a positive real number. 
The first term in Eq.($\ref{D}$) is called a quantum jump term, and the second and third terms in Eq.($\ref{D}$) are continuous non-unitary dissipation terms, respectively. One observes that Eq.($\ref{lindblad}$) can be written as $\text{\cite{Minganti1,MingantiH}}$
\begin{align}
\frac{d\rho}{dt}=-i[H_{\textrm{eff}}\rho-\rho H^{\dagger}_{\textrm{eff}}]+\sum_{i}\Gamma_{i}c_{i}\rho c_{i}^{\dagger},
\end{align}
where we have defined the effective non-Hermitian Hamiltonian $H_{\textrm{eff}}$ by
\begin{align}
H_{\textrm{eff}}=H-i\sum_{i}\Gamma_{i} c^{\dagger}_{i}c_{i}.
\end{align}
If all quantum jump terms are set to 0 by postselection, the time evolution of the system is described by this NHH $\text{\cite{MingantiH}}$.
%\ \ An NHH is said to be PT symmetric $\text{\cite{BenderC.M.Boettcher}}$ if it commutes with the $PT$ operator that combines the parity operator $P$ and the time reversal operator $T$, that is,
%\begin{eqnarray}
%\label{NHHPT}
%[H, PT]=0.
%\end{eqnarray}

% and the region where this approximation holds is called the semiclassical region $\cite{}$.
%\ \ Note that we can extract the process where quantum jumps don't occur using the postselection which is the method to extract specific measurement results $\cite{MingantiH}$. There are some theoretical and experimental papers where the PT symmetry breaking is observed in open quantum systems without quantum jumps $\cite{Li1,Brody,Kawabata,Xiao1,Wu1,Naghiloo}$.

%2.2
\subsection{Eigenvalues and eigenmodes of Liouvillian superoperators}
The Lindblad master equation ($\ref{lindblad}$) is linear in $\rho$, so we can rewrite it with a superoperator, which is a linear operator acting on a vector space of linear operators, as
\begin{align}
\label{120}
\frac{d\rho(t)}{dt}=\hat{\mathcal{L}}\rho(t).
\end{align}
Here $\hat{\mathcal{L}}$ is called the Liouvillian
superoperator. It is known that there is at least one steady state $\rho_{ss}$ if the dimension of the Hilbert space is finite $\text{\cite{ARivas}}$.%Furthermore, it is also known that the steady state is unique under general conditions $\text{\cite{Albert,Nigro}}$. (In particular, in the case of finite spin systems, it is known the steady state is unique if at least one among the Lindblad operators is a ladder operator.) \\

\begin{figure}[h]
   \centering
   \vspace*{-0.2cm}
   \hspace*{-2.5cm}
\includegraphics[bb=0mm 0mm 90mm 150mm,width=0.43\linewidth]{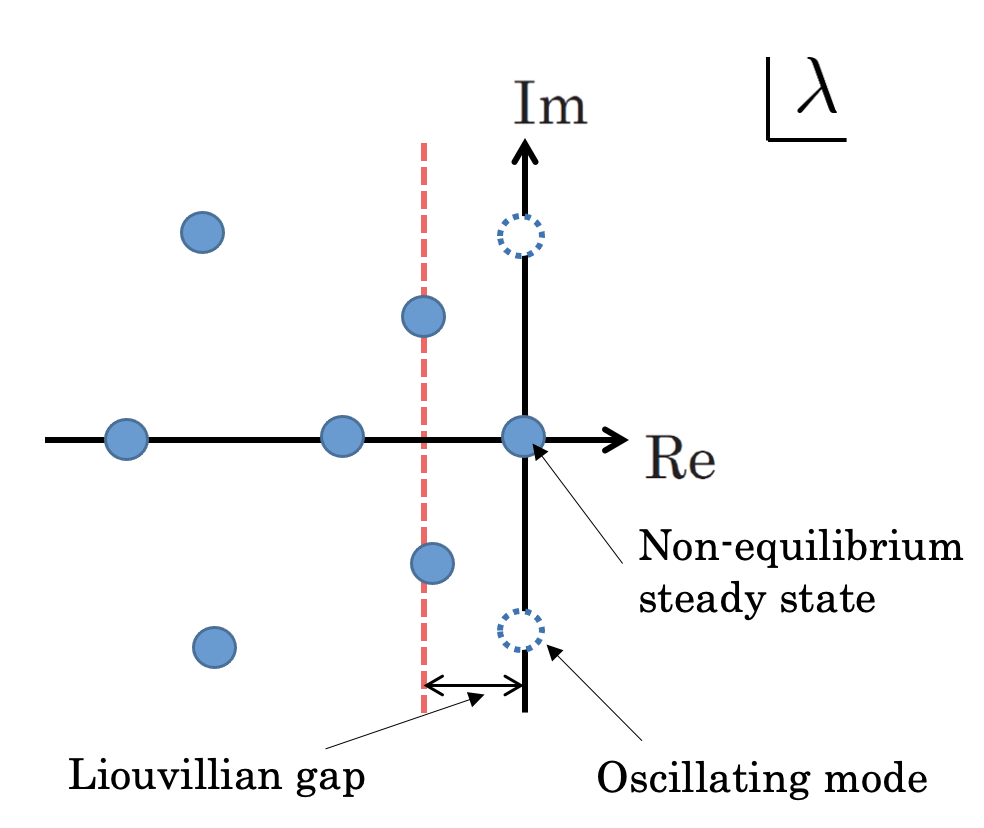}
 \caption{\label{figspectralproperties}Liouvillian spectral properties}
\end{figure}

The eigenvalues $\lambda_{i}$ and the eigenmodes $\rho_{i}$ of the Liouvillian can be obtained by solving the equation,
\begin{align}
\label{1.8}
\hat{\mathcal{L}}\rho_{i}=\lambda_{i}\rho_{i},
\end{align}
where $\rho_{i}$ is not normalized.
It is known that Re[$\lambda_{i}$]$\leq0, \forall i$, and if $\hat{\mathcal{L}}\rho_{i}=\lambda_{i}\rho_{i}$, then $\hat{\mathcal{L}}\rho_{i}^{\dagger}=\lambda_{i}^{*}\rho_{i}^{\dagger}$ $\text{\cite{ARivas}}$. From these properties, the real parts of the Liouvillian eigenvalues are non-positive and the eigenvalue structure is symmetric with respect to the real axis as shown in Fig.$\ref{figspectralproperties}$.

Here, we assume the steady state $\rho_{ss}$ with zero eigenvalue is unique and set up the eigenvalues as $0=|\textrm{Re}[\lambda_{0}]|<|\textrm{Re}[\lambda_{1}]|\leq|\textrm{Re}[\lambda_{2}]|\leq\cdots$. This indicates that $\rho_{ss}=\rho_{0}/\textrm{Tr}[\rho_{0}]$. Also, if the Liouvillian is diagonalizable, the time dependence of the density operators can be written as 
\begin{align}
\label{110}
\rho(t)=\rho_{ss}+\sum_{i\neq0}a_{i}e^{\lambda_{i}t}\rho_{i},
\end{align}
where $a_{i}$ is a time-independent constant $\text{\cite{Minganti2}}$.
This indicates that eigenmodes with eigenvalues which are (very close to) pure imaginary numbers correspond to oscillating terms. Then, the absolute value of the real part of the second maximal eigenvalue $|\textrm{Re}[\lambda_{1}]|$, which is called the Liouvillian gap $\text{\cite{Kessler,Minganti2}}$, is an important physical quantity determining the slowest relaxation rate in the long time limit. Also, it is known that closing the Liouvillian gap is a necessary condition for the occurrence of dissipative phase transitions of the steady state $\text{\cite{Kessler,Minganti2}}$.
%\ \ Next, we consider the points of the parameter space, where some eigenmodes of the Liouvillian coalesce, namely the Liouvillian is nondiagonalizable. These points are called exceptional points $\text{\cite{Heiss}}$. Then, at a Liouvillian exceptional point with order 2 (two eigenmodes coalesce), the time dependence of the density operators can be written $\text{\cite{Arkhipov1}}$ as
%\begin{eqnarray}
%\label{111}
%\rho(t)&=&\rho_{ss}+\sum_{i\neq0}a_{i}e^{\lambda_{i}t}\rho_{i}+a_{\textrm{EP}}\ e^{\lambda_{\textrm{EP}}t}\rho_{\textrm{EP}}\nonumber\\
%&&+a_{\textrm{EP}}^{\prime}\ te^{\lambda_{\textrm{EP}}t}\rho_{\textrm{EP}}^{\prime},
%\end{eqnarray}
%where $a_{\textrm{EP}}$ and $a_{\textrm{EP}}^{\prime}$ are time-independent constants, and $\rho_{\textrm{EP}}$ is an eigenmode with an eigenvalue $\lambda_{\textrm{EP}}$ and $\rho_{\textrm{EP}}^{\prime}$ is a generalized eigenmode. From Eq.($\ref{111}$), we can conjecture that a point where the physical quantities are proportional to time $t$ is a Liouvillian exceptional point with order 2.

%[width=0.4\textwidth,natwidth=500,natheight=410]

%2.3
\subsection{Liouvillian $\mathcal{PT}$ symmetry}
Many studies of Liouvillian $\mathcal{PT}$ symmetry have been conducted recently $\text{\cite{Prosen1,Huber1,Huber2,Prosen2,Huybrechts,Van}}$, but the definition of Liouvillian $\mathcal{PT}$ symmetry has not been established yet. In our arguments, we tentatively adopt (a slightly modified version of ) the proposal of the definition by Huber \textit{et al}. $\text{\cite{Huber2}}$, and discuss its 
consequences and validity.

We say that a Liouvillian associated with the Lindblad equation (\ref{lindblad}) is defined to be $\mathcal{PT}$ symmetric, if the following relation holds,
\begin{align}
\label{HuberPT}
\hat{\mathcal{L}}[\mathbb{PT}(H);\mathbb{PT}^\prime(L_\mu),\mu=1,2,\cdots]=\hat{\mathcal{L}}[H;L_\mu,\mu=1,2,\cdots],\nonumber\\
\\
\label{HuberPT3}\mathbb{PT}(H)=\mathcal{P}\mathcal{T}H(\mathcal{P}\mathcal{T})^{-1}=\mathcal{P}\bar{H}\mathcal{P}^{-1},\ \ \ \ \ \ \ \ \ \ \ \ \ \ \ \ \\
\label{HuberPT2}\mathbb{PT}^\prime(L_{\mu})=\mathcal{P}L_{\mu}^{\dagger}\mathcal{P}^{-1}.\ \ \ \ \ \ \ \ \ \ \ \ \ \ \ \ \ \ \ \ \ \ 
\end{align}
Here $\mathcal{P}$ is a parity operator, $\mathcal{T}$ is the conventional time reversal operator, $i\to -i$, and a dissipation operator $L_{\mu}$ is equal to $\sqrt{\Gamma_{\mu}}c_{\mu}$. The 
parity map $\mathbb{P}$ is the ordinary parity transformation. Note that we have introduced two time reversal maps $\mathbb{T}$ and $\mathbb{T}^{\prime}$. 
The former $\mathbb{T}$ denotes the conventional time reversal transformation and the condition for $H$ in (\ref{HuberPT}) means that a
Hamiltonian satisfies the conventional $\mathcal{PT}$ symmetry ($\ref{NHHPT}$). On the other hand, $\mathbb{T}^{\prime}$ was introduced in $\text{\cite{Huber2}}$ and the condition 
for $L_{\mu}$ in (\ref{HuberPT}) is supposed to represent a physical symmetry under an exchange of gain and loss in terms of dissipation operators. 

In Ref.$\text{\cite{Huber2}}$, the condition for the Hamiltonian part was defined as $\mathbb{PT}^{\prime}(H)=H$. However, we have modified it to $\mathbb{PT}(H)=H$ because, even
though for the two spin model we study there is no difference between the two conditions,   
the latter seems to be  applicable to a broader class of systems. (See the remark after the proof of Theorem 3 below.)
We will still refer to the above criterion of the Liouvillian $\mathcal{PT}$ symmetry as the one by Huber \textit{et al}. 

In particular, for the case of a bipartite system consisting of $A$ and $B$, this definition ($\ref{HuberPT}$) is written as 
\begin{align}
\label{two}
\hat{\mathcal{L}}[\mathbb{PT}(H);\mathbb{PT}^{\prime}(L_A),\mathbb{PT}^{\prime}(L_B)]&=\hat{\mathcal{L}}[H; L_A, L_B],
\end{align}
where we define the parity operator as the space reversal operator,
\begin{align}
\mathcal{P}(A\otimes B)\mathcal{P}^{-1}&=B\otimes A.
\end{align}
 We remark that this definition differs from Prosen's proposal of the definition of $\mathcal{PT}$ symmetric Liouvillian superoperators introduced in Ref.$\text{\cite{Prosen1}}$. 

\subsection{Bosonic third quantization}
We review the framework of the bosonic third quantization which is a general method to solve the Lindblad equation for open quadratic systems with linear bath operators $\text{\cite{Prosen4}}$. We will explain the general case in Appendix A.

First, we rewrite the Liouvillian superoperator in Eq.($\ref{120}$) as
\begin{eqnarray}
\label{lio1}
\hat{\mathcal{L}}&=&-i\hat{H}^{L}+i\hat{H}^{R}\nonumber\\
&+&\sum_{\mu}2{\hat{L}_{\mu}}^{L}{\hat{L}^{\dagger R}_{\mu}}-{\hat{L}_{\mu}^{\dagger L}}{\hat{L}_{\mu}}^{L}-{\hat{L}_{\mu}}^{R}{\hat{L}_{\mu}^{\dagger R}},
\end{eqnarray}
where we define the right superoperator and the left superoperator as
\begin{eqnarray}
\hat{O}^{R}\rho=\rho O,\ \ \ \ \hat{O}^{L}\rho=O\rho,
\end{eqnarray}
respectively.

The Hamiltonian and dissipation operators for an arbitrary quadratic system of $n$ bosons
with linear bath operators can be written as
\begin{eqnarray}
\label{Hamiltonian11}
H&=&\underline{a}^{\dagger}\cdot\textbf{H}\underline{a}+\underline{a}\cdot\textbf{K}\underline{a}+\underline{a}^{\dagger}\cdot\bar{\textbf{K}}\underline{a}^{\dagger},\\
\label{dissipation11}L_{\mu}&=&\underline{l}_{\mu}\cdot\underline{a}+\underline{k}_{\mu}\cdot\underline{a}^{\dagger}.
\end{eqnarray}
where $\textbf{H}=\textbf{H}^{\dagger}$ and $\textbf{K}=\textbf{K}^{T}$ are $n\times n$ matrices, $\underline{a}=(a_{1},a_{2},...,a_{n})^{T}$ is a vector which consists of annihilation bosonic operators, $\underline{a}^{\dagger}=(a_{1}^{\dagger},a_{2}^{\dagger},...,a_{n}^{\dagger})^{T}$ is a vector which consists of creation bosonic operators, and $\underline{l}_{\mu}$ and $\underline{k}_{\mu}$ are $n$ vectors for dissipation rates. Here the underline means a vector. 
Also, we define the set of $4n$ maps $\hat{a}_{\nu,j}$, and $\hat{a}_{\nu,j}^{\prime}$,
\begin{align}
\hat{a}_{0,j}=\hat{a}^{L}_{j},\ \ \ \ \hat{a}^{\prime}_{0,j}=\hat{a^{\dagger}}^{L}_{j}-\hat{a^{\dagger}}^{R}_{j},\nonumber\\
\label{511}\hat{a}_{1,j}=\hat{a^{\dagger}}^{R}_{j},\ \ \ \ \hat{a}^{\prime}_{1,j}=\hat{a}^{R}_{j}-\hat{a}^{L}_{j}\ \ 
\end{align}
with 
\begin{equation}
\label{CCR}
[\hat{a}_{\nu,j},\hat{a}^{\prime}_{\mu,k}]=\delta_{\nu,j}\delta_{\mu,k}, \ \ \ \  [\hat{a}_{\nu,j},\hat{a}_{\mu,k}]=[\hat{a}^{\prime}_{\nu,j},\hat{a}^{\prime}_{\mu,k}]=0,
\end{equation}
where $\ j,k=1,...,n,\ \nu,\mu=0,1.$ Now, we define $\underline{\hat{b}}=(\underline{\hat{a}},\underline{\hat{a}}^{\prime})^{T}=(\underline{\hat{a}}_{0},\underline{\hat{a}}_{1},\underline{\hat{a}}^{\prime}_{0},\underline{\hat{a}}^{\prime}_{1})^{T}$ as a $4n$ vector with ($\rm\ref{511}$) and then rewrite the Liouvillian
($\rm\ref{lio1}$) in a symmetric form
\begin{equation}
\label{161}
\hat{\mathcal{L}}=\underline{\hat{b}}\cdot\textbf{S}\underline{\hat{b}}-S_{0}\hat{\1},
\end{equation}
where $\textbf{S}$ is a complex symmetric $4n \times 4n$ matrix which can be written in
terms of two $2n \times 2n$ matrices $\textbf{X}$ and $\textbf{Y}$ as 
\begin{equation}
\label{S1}
\textbf{S}=\left[
   \begin{array}{cc}
     \textbf{0} & - \textbf{X} \\
      -\textbf{X}^{T}  & \textbf{Y}
   \end{array}
  \right],
\end{equation}
where
\begin{eqnarray}
\label{X111}
\textbf{X}:=\frac{1}{2}\left(
   \begin{array}{cc}
     i\bar{\textbf{H}}-\bar{\textbf{N}}+\textbf{M} & -2i\textbf{K}-\textbf{L}+\textbf{L}^{T}  \\
      2i\bar{\textbf{K}} -\bar{\textbf{L}}+\bar{\textbf{L}}^{T} & -i\textbf{H}-\textbf{N}+\bar{\textbf{M}}
   \end{array}
  \right),
\end{eqnarray}
and
\begin{eqnarray}
\label{Y111}
\textbf{Y}:=\frac{1}{2}\left(
   \begin{array}{cc}
     -2i\bar{\textbf{K}}-\bar{\textbf{L}}-\bar{\textbf{L}}^{T}  & 2\textbf{N} \\
      2\textbf{N}^{T} & 2i\textbf{K}-\textbf{L}-\textbf{L}^{T} 
   \end{array}
  \right).
\end{eqnarray}
Here $S_{0}=$tr$\textbf{X}$ and the matrices $\textbf{M}$, $\textbf{N}$ and $\textbf{L}$ are defined as
\begin{eqnarray}
\label{M}\textbf{M}&:=&\sum_{\mu}\underline{l}_{\mu}\otimes\underline{\bar{l}}_{\mu}=\textbf{M}^{\dagger},\\ 
\label{N}\textbf{N}&:=&\sum_{\mu}\underline{k}_{\mu}\otimes\underline{\bar{k}}_{\mu}=\textbf{N}^{\dagger},\\ 
\label{L}\textbf{L}&:=&\sum_{\mu}\underline{l}_{\mu}\otimes\underline{\bar{k}}_{\mu}.
\end{eqnarray}

Next, we assume that the matrix $\textbf{X}$ ($\rm\ref{X111}$) is diagonalizable, that is, it can be written as
\begin{eqnarray}
\label{20}
\textbf{X}=\textbf{P}{\boldsymbol{\Delta}}\textbf{P}^{-1},\ \ \ \ \ {\boldsymbol{\Delta}}=\textrm{diag}\{\beta_{1},\beta_{2},...,\beta_{2n} \},
\end{eqnarray}
where $\textbf{P}$ is a $2n$ $\times$ $2n$ matrix.
Then, if the real parts of all the eigenvalues $\beta$ are positive, it is shown that the Liouvillian ($\rm\ref{lio1}$) can be written as 
\begin{eqnarray}
\label{mukatuku1}
\hat{\mathcal{L}}=-2\sum_{r=1}^{2n}\beta_{r}\hat{\zeta}^{\prime}_{r}\hat{\zeta}_{r},
\end{eqnarray}
where we define the $4n$ normal master-mode maps $\hat{\underline{\zeta}}$, and $\hat{\underline{\zeta}}^{\prime}$ as 
\begin{eqnarray}
\label{241}
(\hat{\underline{\zeta}},\hat{\underline{\zeta}}^{\prime})^{T}:=(\textbf{P}^{T}(\hat{\underline{a}}-\textbf{Z}\hat{\underline{a}}^{\prime}),\textbf{P}^{-1}\hat{\underline{a}}^{\prime})^T
\end{eqnarray}
with
\begin{eqnarray}
\label{zeta11}
[\hat{\zeta}_{r},\hat{\zeta}^{\prime}_{s}]=\delta_{r,s},\ \ \ \ \ [\hat{\zeta}_{r},\hat{\zeta}_{s}]=[\hat{\zeta}_{r}^{\prime},\hat{\zeta}^{\prime}_{s}]=0.
\end{eqnarray}
Also, it is known that the matrix $\textbf{Z}$ in Eq.($\rm\ref{241}$) can be obtained from
\begin{eqnarray}
\label{Lyapunov}
\textbf{X}^{T}\textbf{Z}+\textbf{Z}\textbf{X}=\textbf{Y}.
\end{eqnarray}
From Eqs.($\rm\ref{mukatuku1}$), ($\rm\ref{zeta11}$), the following theorems can be obtained $\text{\cite{Prosen4}}$.

\textbf{\textit{Theorem 1}}: If the matrix $\textbf{X}$ ($\rm\ref{X111}$) is diagonalizable and the real parts of all the eigenvalues are positive, i.e. $\forall j,\rm{Re} [\beta_{j}]>0$, the unique non-equilibrium steady state $\rho_{ss}$ exists, satisfying
\begin{eqnarray}
\label{327}
\hat{\mathcal{L}}\rho_{ss}=0.
\end{eqnarray}

\textbf{\textit{Theorem 2}}: If the matrix $\textbf{X}$ ($\rm\ref{X111}$) is diagonalizable and the real parts of all the eigenvalues are positive, i.e. $\forall j,\rm{Re} [\beta_{j}]>0$, the full spectrum of the Liouvillian is given by a $2n$ component multi-index of super-quantum numbers $\underline{m}\in\mathbb{Z}_{+}^{2n}$,
\begin{eqnarray}
\label{newbasis}
\hat{\mathcal{L}}\rho_{\underline{m}}=\lambda_{\underline{m}}\rho_{\underline{m}},\ \ \lambda_{\underline{m}}=-2\sum_{r=1}^{2n}m_{r}\beta_{r},\ \ \rho_{\underline{m}}=\prod_{r=1}^{2n}\frac{(\hat{\zeta}^{\prime}_{r})^{m_{r}}}{\sqrt{m_{r}!}}\rho_{ss}.\nonumber\\
\end{eqnarray}
These theorems indicate that an exact eigenvalue structure of the Liouvillian can be obtained if the real parts of all the eigenvalues $\beta$ are positive.

%3
\section{Eigenvalue structure and dynamics of an open 2-spin model}
\subsection{Open 2-spin-$S$ model with criterion of Huber \textit{et al}. of Liouvillian $\mathcal{PT}$ symmetry}
We consider a Hamiltonian of an open 2 spin-$S$ system with $XX$ interaction as
\begin{align}
\label{2spindefinition}
H=\frac{g}{2S}(S_{A}^{+}S_{B}^{-}+ \textrm{H.c.}).
\end{align}
Also, we consider alternate pumping of spins along opposite directions as dissipations. Then, the time evolution is described by Lindblad equation as
\begin{align}
\label{2spindefinition2}
\frac{\partial}{\partial t}\rho=\hat{\mathcal{L}}\rho=-i[H,\rho]+\frac{\Gamma_{g}}{2S}\mathcal{D}[\hat{S}_A^+]\rho+\frac{\Gamma_{l}}{2S}\mathcal{D}[\hat{S}_B^-]\rho,
\end{align}
where $S^{\pm}=S^{x}\pm iS^{y}$. Here, $g$ is the strength of coupling, and $\Gamma_{g}$ and $\Gamma_{l}$ are the strengths of dissipations. This model satisfies the criterion of Huber et al. of Liouvillian $\mathcal{PT}$ symmetry ($\ref{HuberPT}$) when $\Gamma_{g}=\Gamma_{l}$. [Note that this model satisfies ($\ref{HuberPT}$) even if the definition of Hamiltonian $\mathcal{PT}$ symmetry ($\ref{HuberPT3}$) is written as $\mathbb{PT}^{\prime}(H)=H$, not $\mathbb{PT}(H)=H$, since $H$ ($\ref{2spindefinition}$) is real symmetric.]

\begin{figure}[htb]
  \begin{center}
  \vspace*{-0.5cm}
     \hspace*{-6.8cm}
\includegraphics[bb=0mm 0mm 90mm 150mm,width=0.41\linewidth]{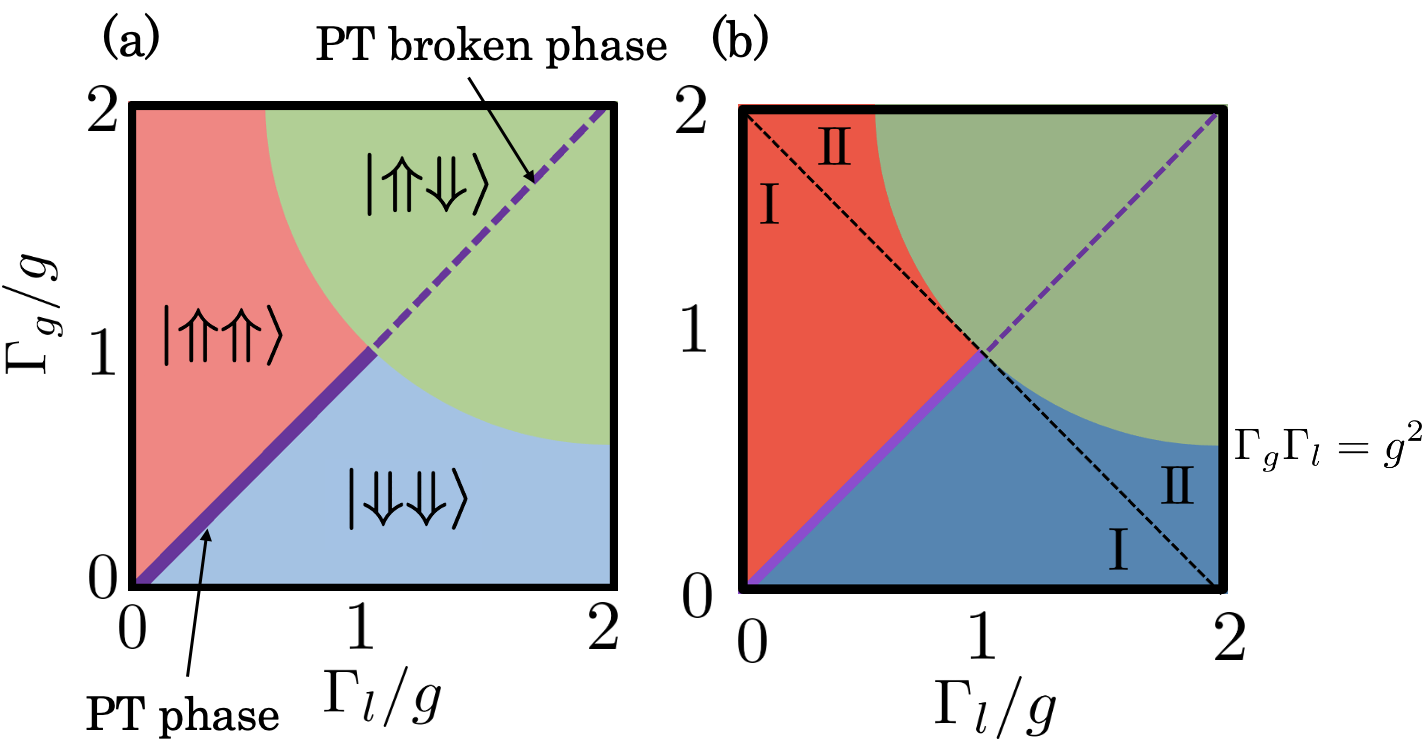}
    \caption{Phase diagrams of the open 2-spin model for $S\gg1$. (a) Phase diagram determined from the normalized magnetizations (and the symmetry parameter) of the steady state. The green region with $\Gamma_{g}\Gamma_{l}>g^{2}$ is called the AFM $\ket{\Uparrow\Downarrow}$ phase, the red (\textit{resp}. blue) one with $\Gamma_{g}\Gamma_{l}<g^{2}$ and $\Gamma_{g}>\Gamma_{l}$ (\textit{resp}. $\Gamma_{g}<\Gamma_{l}$) is called the FM $\ket{\Uparrow\Uparrow}$ phase (\textit{resp}. the FM $\ket{\Downarrow\Downarrow}$ phase). Also, by the behavior of the symmetry parameter, the purple line with $\Gamma_{g}\Gamma_{l}<g^{2}$ and $\Gamma_{g}=\Gamma_{l}$ can be regarded as the $\mathcal{PT}$ phase, and the AFM phase for $\Gamma_{g}=\Gamma_{l}$ (dashed purple line) and the point $\Gamma_{g}=\Gamma_{l}=g$ can be regarded as the $\mathcal{PT}$ broken phase and the $\mathcal{PT}$ symmetry breaking point, respectively. (b) Phase diagram determined from the relaxation time (dynamics). The relaxation time diverges at phase boundary $\Gamma_{g}\Gamma_{l}=g^{2}$ and $\Gamma_{g}=\Gamma_{l}<g$. In particular, the physical quantities such as the magnetization oscillate in the long time limit in the $\mathcal{PT}$ phase while they decay without oscillation in the $\mathcal{PT}$ broken phase. Moreover, the FM phase can be divided by the condition of whether the physical quantities such as the magnetization oscillate or not. We call the FM phase with $\Gamma_{g}+\Gamma_{l}<2g$ FM phase I and the FM phase with $\Gamma_{g}+\Gamma_{l}>2g$ FM phase I$\hspace{-.1em}$I.
 }
    \label{phasedia}
  \end{center}
\end{figure}

For this model, there are several phases with a large $S$ as shown in Fig.\ref{phasedia} (a) because of the competition of dissipations and the interaction $\text{\cite{Huber1,Huber2}}$. Each phase has been determined by the normalized magnetizations $\braket{M^{z}_{A}}:=\braket{S^{z}_{A}}/S$ and $\braket{M^{z}_{B}}:=\braket{S^{z}_{B}}/S$ for $S\gg1$. The region with $\Gamma_{g}\Gamma_{l}>g^{2}$ is called the anti-ferromagnetic (AFM) $\ket{\Uparrow\Downarrow}$ phase, where $\braket{M^{z}_{A}}=-\braket{M^{z}_{B}}=1$. The one with $\Gamma_{g}\Gamma_{l}<g^{2}$ and $\Gamma_{g}>\Gamma_{l}$ (\textit{resp}. $\Gamma_{g}<\Gamma_{l}$) is called the ferromagnetic (FM) $\ket{\Uparrow\Uparrow}$ phase (\textit{resp}. the FM  $\ket{\Downarrow\Downarrow}$ phase), where $\braket{M^{z}_{A}}=\braket{M^{z}_{B}}=1$ (\textit{resp}. $\braket{M^{z}_{A}}=\braket{M^{z}_{B}}=-1$). On the line segment with $\Gamma_{g}\Gamma_{l}<g^{2}$ and $\Gamma_{g}=\Gamma_{l}$, the normalized magnetizations of the steady state can not be determined for a general initial condition. But it is known that this model even on the segment has a unique steady state when $S$ is finite $\text{\cite{Nigro}}$ and the normalized magnetizations of this eigenmode (steady state) numerically approach zero as $S$ increases $\text{\cite{Huber1}}$. Therefore it can be distinguished from other phases by this property. %From dynamical point of view this line segment will be the PT phase, as we will show later. 

%And the one with $\Gamma_{g}\Gamma_{l}<g^{2}$ and $\Gamma_{g}=\Gamma_{l}$ is called the PT phase. Here, (as we will show later) the PT phase is \textit{the dynamical phase} and the AM and FM phases are the stationary phases when $S$ is infinite. So, the normalized magnetizations of the steady state can not be determined for a general initial condition in the PT phase. On the other hand, when $S$ is finite, it is known that this model has the unique steady state $\text{\cite{Nigro}}$. Also, it is numerically shown that the normalized magnetizations of this eigenmode (steady state) approach zero in the PT phase as $S$ increases $\text{\cite{Huber1}}$, and then the PT phase can be distinguished from other phases by its stationary nature.

%Here, the normalized magnetization is determined in the PT phase by using an eigenmode with 0 eigenvalue since the PT phase is not relaxed when S is infinite.

%choosing an eigenmode with 0 eigenvalue which is also the steady state when S is finite. 

%Here, (we will show later that) the PT phase is \textit{the dynamical phase} and the AM and FM phases are the stationary phases.

There are two results suggesting that definition of Huber \textit{et al}. ($\ref{HuberPT}$) is a good definition of Liouvillian $\mathcal{PT}$ symmetry. The first is the behavior of the symmetry parameter $\text{\cite{Kepesidis}}$, which provides a measure for the parity symmetry of the density operator, when $\Gamma_{g}=\Gamma_{l}=\Gamma$, namely the model is $\mathcal{PT}$ symmetric according to the definition of Huber \textit{et al}. In particular, it changes from 0 to a finite value at $\Gamma=g$ for a large $S$. This implies that the $\mathcal{PT}$ symmetry breaking occurs at $\Gamma=g$ $\text{\cite{Huber2}}$. Therefore, the segment with $\Gamma_{g}\Gamma_{l}<g^{2}$ and $\Gamma_{g}=\Gamma_{l}$ and the AFM phase for $\Gamma_{g}=\Gamma_{l}$ can be regarded as the $\mathcal{PT}$ phase and the $\mathcal{PT}$ broken phase, respectively. Then, the point $\Gamma_{g}=\Gamma_{l}=g$ can also be regarded as the $\mathcal{PT}$ symmetry breaking point. Furthermore, it has been shown that other physical quantities, such as purity and negativity, clearly change at $\Gamma=g$ for a large $S$ $\text{\cite{Huber2}}$. Here, these physical quantities in the $\mathcal{PT}$ phase are determined by the limit from both FM phases. Furthermore, numerical calculations show that these physical quantities of this eigenmode (steady state) approach the same values as $S$ increases in the $\mathcal{PT}$ phase $\text{\cite{Huber1,Huber2}}$.

The second result is a numerical analysis of the dynamics of the magnetization with a finite $S$ $\text{\cite{Huber2}}$. This shows that the time evolution of the magnetization undergoes a transition from an oscillatory to an overdamped behavior at the $\mathcal{PT}$ symmetry breaking point according to the definition of Huber \textit{et al}. only for a finite time.

However, it has been observed that the magnetizations decay after a long time even in the $\mathcal{PT}$ phase because the state decays toward the steady state when the system's dimension is finite and dissipation operators are spin ladder operators $\text{\cite{Nigro}}$. Moreover, the eigenvalue structure does not clearly change above and below the $\mathcal{PT}$ symmetry breaking point for a finite $S$.
In the $\mathcal{PT}$ symmetric Hamiltonian case, the eigenvalues and the dynamics in the long time limit also show a clear change at the $\mathcal{PT}$ symmetry breaking point $\text{\cite{Ramezani}}$. Therefore, it has not been clear if one can call the transition in the open 2-spin model a Liouvillian $\mathcal{PT}$ phase transition. 

Also, once the eigenvalue structure is obtained, it is very easy to check the order of degeneracies and exceptional points, the closure of Liouvillian gaps, and the presence of oscillations in the long time limit (dynamical phases), which have important physical meanings.

%[width=0.4\textwidth,natwidth=590,natheight=460]

\subsection{Analysis of eigenvalue structure}
In this section, we investigate the eigenvalue structure and the dynamics of the magnetization for the open 2-spin model with the third quantization $\text{\cite{Prosen3,Prosen4}}$ explained in section I\hspace{-1pt}I. E.
It can not be applied to spin systems directly, but it becomes possible for spins with large total spin, after using the Holstein-Primakoff (HP) approximation $\text{\cite{Holstein}}$. In this paper, we only deal with quadratic systems, i.e. we focus on the region where $N/2S\ll$1 with the photon number $N=\braket{c^{\dagger}c}\sim\mathcal{O}(1)$. If this condition  $N/2S\ll$1 is violated, the nonlinear terms will not be neglected and then we must consider significant effects such as nonlinear bosonic saturation.

The third quantization for bosonic systems has not been widely discussed, but it is an effective method to obtain not only the eigenvalues and the eigenmodes but also the time evolution of the physical quantities. We explain the HP approximation Appendix B.

In the context of the third quantization, a matrix $\textbf{X}$, which constitutes the Liouvillian, plays an important role. In this section, we do not give details of our calculations, but state only the results. For the derivation, see Appendix C.
In section IV, we will discuss relations between the physical quantities and the criterion of Huber \textit{et al}. of Liouvillian $\mathcal{PT}$ symmetry for general quadratic bosonic systems using the third quantization. 

\begin{figure}[h]
  \begin{center}
  \vspace*{1.6cm}
     \hspace*{-6.3cm}
\includegraphics[bb=0mm 0mm 90mm 150mm,width=0.41\linewidth]{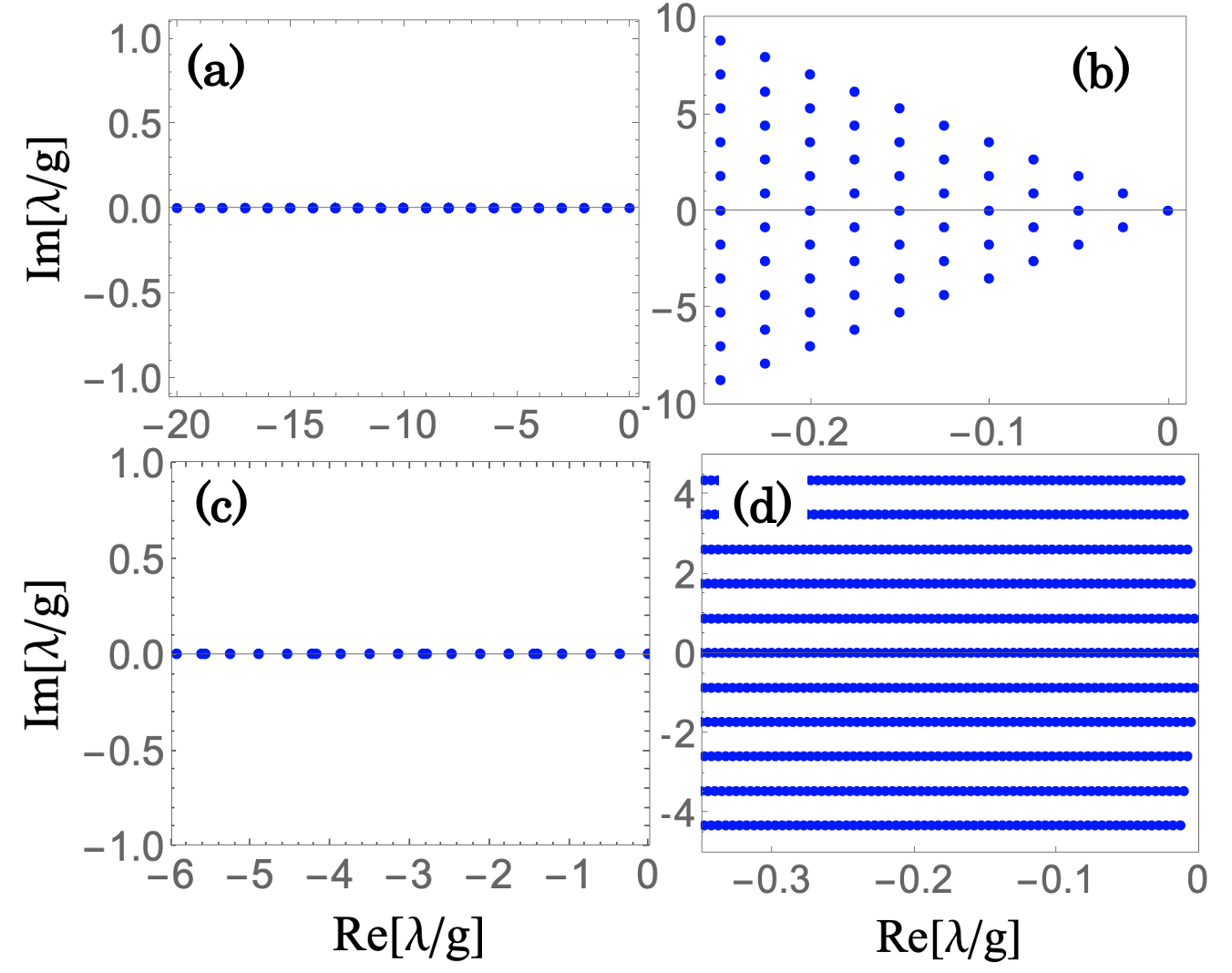}
    \caption{Eigenvalue structure of the open 2-spin model. We set parameters as (a) $\Gamma_{g}=\Gamma_{l}=2$, and $g=1$ in the AM ($\mathcal{PT}$ broken) phase (b) $\Gamma_{g}=0.5, \Gamma_{l}=0.45$, and $g=1$ in the FM phase I (c) $\Gamma_{g}=2, \Gamma_{l}=0.25$, and $g=1$ in FM phase I$\hspace{-.1em}$I (d) $\Gamma_{g}=0.5, \Gamma_{l}=0.495$, and $g=1$ in FM phase I (in the vicinity of the $\mathcal{PT}$ phase). %In particular, all the eigenvalues are real in the AM (PT broken) phase, while some eigenvalues are near pure imaginary numbers in the near PT phase.
    }
    \label{AMeigenvalue}
  \end{center}
\end{figure}

\begin{figure}[h]
  \begin{center}
  \vspace*{1.3cm}
  \hspace*{-5.4cm}
\includegraphics[bb=0mm 0mm 90mm 150mm,width=0.41\linewidth]{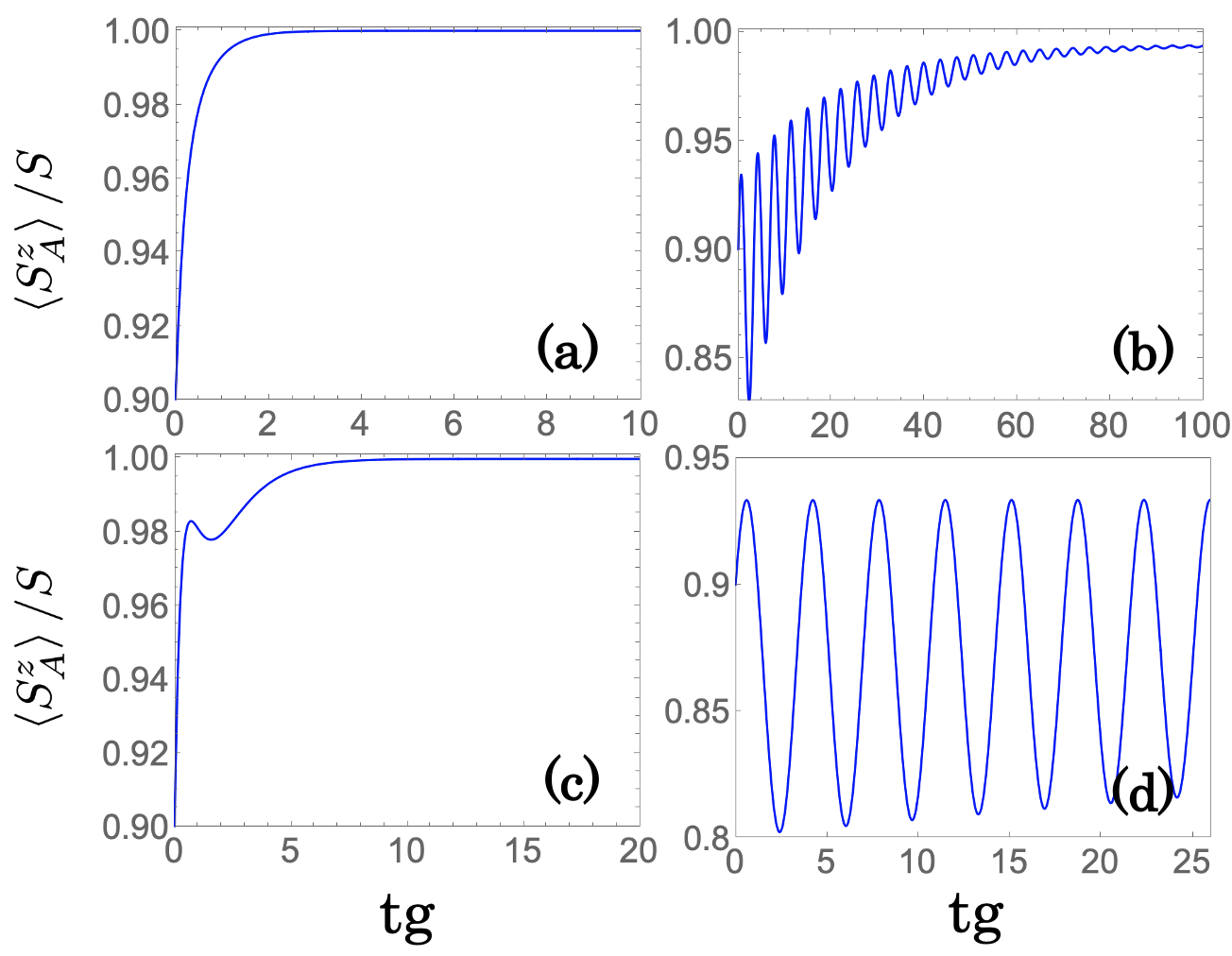}
    \caption{Time evolution of the normalized magnetization $\braket{S_{A}^{z}}(t)/S$ for the open 2-spin model. We set the parameters as (a) $\Gamma_{g}=\Gamma_{l}=2$, and $g=1$ in AFM (PT broken) phase (b) $\Gamma_{g}=0.5, \Gamma_{l}=0.45$, and $g=1$ in FM phase I (c) $\Gamma_{g}=2, \Gamma_{l}=0.25$, and $g=1$ in FM phase I$\hspace{-.1em}$I (d) $\Gamma_{g}=0.5, \Gamma_{l}=0.495$, and $g=1$ in the FM phase I (in the vicinity of the $\mathcal{PT}$ phase) with $S=1000$, and $\braket{S_{A}^{z}}(0)=900$. Note that we use the expression of the time dependence of the two point functions ($\ref{Zt}$), which will be derived in Appendix E.}
    \label{dynamics}
  \end{center}
\end{figure}

 In the AFM phase, we obtain all the Liouvillian eigenvalues $\lambda^{AFM}$ for $S\gg1$ by combining Theorem 2 and calculations in Appendix C.1 as
\begin{eqnarray}
\label{lambdaAM}
\lambda^{AFM}=-2[(m_{1}+m_{2})\beta^{AFM}_{+}+(m_{3}+m_{4})\beta^{AFM}_{-}],
\end{eqnarray}
where $m_{i}\in\mathbb{Z}_{+}$ ($i$=1, 2, 3, 4) and
\begin{eqnarray}
\label{betaAM1}
\beta^{AFM}_{\pm}=\frac{1}{4}\left(\Gamma_g+\Gamma_l\pm\sqrt{(\Gamma_g-\Gamma_l)^2+4g^2}\right)>0.
\end{eqnarray}
Here $\beta^{AFM}_{\pm}$ are the eigenvalues of the matrix $\textbf{X}$ ($\rm\ref{XX}$) in Appendix C and $\beta^{AFM}_{\pm}$ are real.
We plot eigenvalues of the AFM ($\mathcal{PT}$ broken) phase in Fig.$\ref{AMeigenvalue}$ (a). We can find that all the eigenvalues are real and then the state decays toward the steady state without oscillation. In fact, the normalized magnetization decays exponentially toward the steady state as shown in Fig.$\ref{dynamics}$ (a). The time evolution of the magnetization can be obtained from the expression of the time dependence of the two point functions for a more general system in Eq.($\ref{Zt}$) below, which will be derived in Appendix E. Also, we can show that Liouvillian gap is closed at the phase boundary $\Gamma_{g}\Gamma_{l}=g^{2}$ in Eq.($\rm\ref{limit1}$).

%[width=0.63\textwidth,natwidth=740,natheight=500]

In the FM phase, we obtain all the Liouvillian eigenvalues $\lambda^{FM}$ for $S\gg1$ with Theorem 2, A1 in Appendix A and calculations in Appendix C.2 as
\begin{eqnarray}
\label{lambdaFM}
\lambda^{FM}=-2[(m_{1}+m_{2})\beta^{FM}_{+}+(m_{3}+m_{4})\beta^{FM}_{-}],
\end{eqnarray}
where $m_{i}\in\mathbb{Z}_{+}$ ($i$=1, 2, 3, 4) and
\begin{eqnarray}
\label{betaFM2}
\beta^{FM}_{\pm}=\frac{1}{4}\left(|\Gamma_g-\Gamma_l|\pm\sqrt{(\Gamma_g+\Gamma_l)^2-4g^2}\right).
\end{eqnarray}
Here $\beta^{FM}_{\pm}$ in the FM $\ket{\Uparrow\Uparrow}$ phase are eigenvalues of the matrix $\textbf{X}$ [Eq.($\rm\ref{FMX}$)] and Re[$\beta^{FM}_{\pm}]>0$.
We plot eigenvalues of the FM phase in Figs.$\ref{AMeigenvalue}$(b), and $\ref{AMeigenvalue}$(c).

From Eq.(\ref{betaFM2}), we can find that the eigenvalue structure changes at the line, $\Gamma_{g}+\Gamma_{l}=2g$. In fact, the normalized magnetization decays with oscillation if $\Gamma_{g}+\Gamma_{l}<2g$ (Fig.$\ref{dynamics}$ (b)) and it decays without oscillation if $\Gamma_{g}+\Gamma_{l}>2g$ (Fig.$\ref{dynamics}$ (c)). Hereafter, we call the region with $\Gamma_{g}+\Gamma_{l}<2g$ the FM phase I and the region with $\Gamma_{g}+\Gamma_{l}>2g$ the FM phase I$\hspace{-.1em}$I respectively. Note that this transition is not a dissipative phase transition of the steady state because the Liouvillian gap is not closed. Moreover, we can say that all the points on the line $\Gamma_{g}+\Gamma_{l}=2g$ are Liouvillian exceptional points $\text{\cite{Heiss,Kanki}}$, which are the points of the parameter space where some eigenmodes of the Liouvillian coalesce, since the matrix $\textbf{X}$ ($\rm\ref{FMX}$) is nondiagonalizable on this line, (see the discussion after Theorem A1 in Appendix A.) The same exceptional point line 
was found in $\text{\cite{Roccati1}}$ and a similar one for a similar model was observed in $\text{\cite{Arkhipov1}}$. 
If we further consider a limit to the $\mathcal{PT}$ symmetry breaking point, $\Gamma_g=\Gamma_l=g$, from the FM phases, the matrix $\textbf{X}$ ($\rm\ref{FMX}$) 
approaches a nondiagonalizable matrix. This suggests that the $\mathcal{PT}$ symmetry breaking point is a Liouvillian exceptional point, though one has to consider also the limit from the AFM phase to have a definite conclusion. 

Next we discuss the eigenvalues at the phase boundary $\Gamma_{g}=\Gamma_{l}(=\Gamma)$ on which criterion of Huber \textit{et al}. holds. 
Taking the limit $|\Gamma_{g}-\Gamma_{l}|\to 0$ in 
the FM phases, we find
\begin{eqnarray}
\label{PTlimit}
\beta^{FM}_{\pm}\to\pm\frac{i}{2}\sqrt{g^{2}-\Gamma^{2}}. 
\end{eqnarray}
This shows that the Liouvillian gap is closed at the phase boundary, $\Gamma_{g}=\Gamma_{l}$, and there exist some pure imaginary eigenvalues in the $\mathcal{PT}$ phase. Also, $m_{i}$ can be arbitrarily large, so we can obtain the eigenvalue structure of 
the $\mathcal{PT}$ phase. %as shown in Fig.$\ref{AMeigenvalue}$ (d). 
Since some eigenvalues are pure imaginary numbers, the physical quantities oscillate even in the long time limit. In other words, this shows that the $\mathcal{PT}$ phase is \textit{the dynamical phase}. In fact, the relaxation time of the normalized magnetization in the vicinity of the $\mathcal{PT}$ phase is very long as shown in Fig.$\ref{dynamics}$ (d). Moreover, it is exactly found from Eqs.($\ref{lambdaFM}$), ($\ref{PTlimit}$) that an infinite number of eigenvalues approach 0 in the $\mathcal{PT}$ phase although it was expected numerically in Ref.$\text{\cite{Huber1}}$. This result (the existence of infinite 0 eigenvalues) can not be obtained from the evolution matrix of the first-order moment.

The phase transition from the FM $\ket{\Uparrow\Uparrow}$ phase to the FM $\ket{\Downarrow\Downarrow}$ phase is a dissipative first-order phase transition when determined from the discontinuous behavior of magnetization. Furthermore, it is known that the steady state at the transition point ($\mathcal{PT}$ phase) is a highly mixed state, where the purity Tr$[\rho^{2}]$ is almost 0 (e.g. uniform distribution $\rho\propto\1$) $\text{\cite{Huber1,Huber2}}$. However, in the general theory of dissipative first-order phase transitions, the steady state at the transition point is expected to the equiprobable mixture of the opposite FM phases, namely $\rho=(\ket{\Uparrow\Uparrow}\bra{\Uparrow\Uparrow}+\ket{\Downarrow\Downarrow}\bra{\Downarrow\Downarrow})/2$ $\text{\cite{Minganti2}}$.

\begin{figure}[htb]
  \begin{center}
  \vspace*{-1.2cm}
  \hspace*{-5.7cm}
\includegraphics[bb=0mm 0mm 90mm 150mm,width=0.40\linewidth]{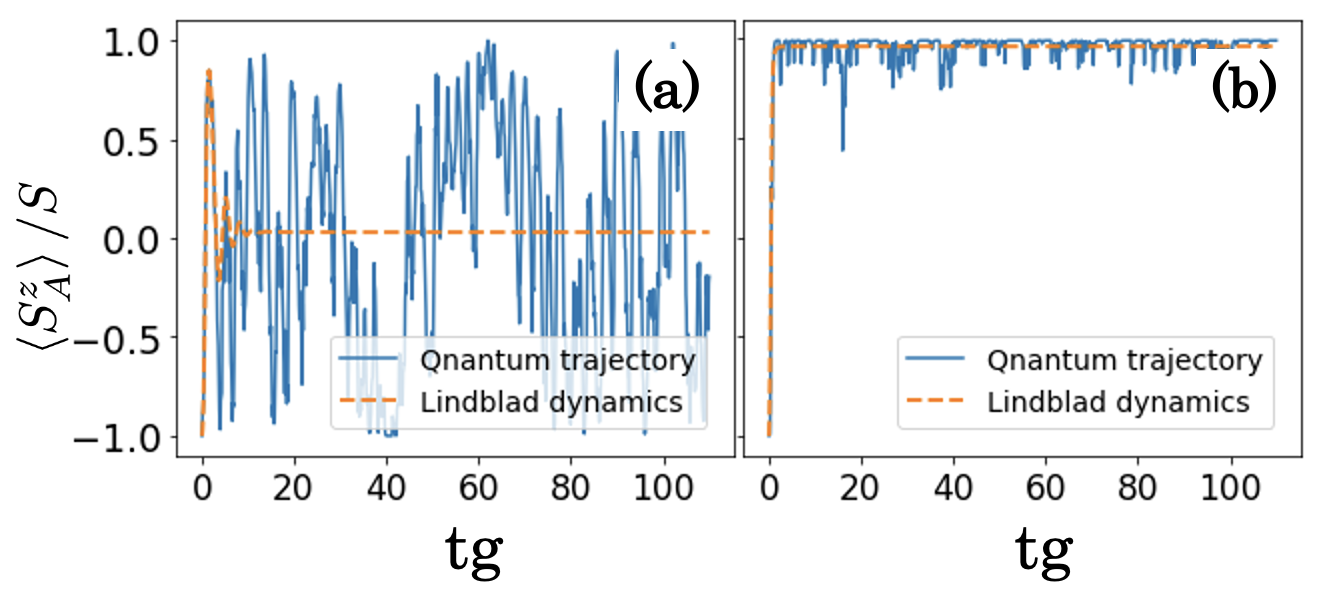}
    \caption{Normalized magnetization $\braket{S_{A}^{z}}(t)/S$ for a single quantum trajectory (a Monte-Carlo trajectory) (blue solid line) and the Lindblad dynamics (orange dashed line) for $S=10$. We set the parameters as (a) $\Gamma_{g}=\Gamma_{l}=\Gamma=0.5, g=1$ in the $\mathcal{PT}$ phase, and  (b) $\Gamma_{g}=\Gamma_{l}=\Gamma=1.5, g=1$ in the $\mathcal{PT}$ broken phase, where the initial state is set as $\rho(t=0)=\ket{-S}\bra{-S}\otimes \ket{S}\bra{S}$. Here, we have used QuTiP $\text{\cite{Johansson1,Johansson2}}$ to numerically obtain quantum trajectories and Lindblad dynamics.}
    \label{tra}
  \end{center}
\end{figure}

To examine the reason for this, we have performed the quantum trajectory analysis (the Monte-Carlo trajectory simulation) $\text{\cite{Daley,Johansson1}}$ for a finite $S$.
In the PT phase, the quantum trajectory analysis indicates that the quantum fluctuation is very large (i.e., the normalized magnetization takes from -1 to 1) and  Lindblad dynamics analysis indicates that the normalized magnetization is 0 in the long time limit (Fig.$\ref{tra}$ (a)). This implies that a highly mixed steady state with zero magnetization mean is realized for the Lindblad dynamics since it is equivalent to the averaged dynamics over various quantum trajectories. 
Physically speaking, in the $\mathcal{PT}$ phase, the contribution of non unitary continuous dissipations (gain and loss) are canceled out due to sufficient interactions, but the contribution of quantum jumps remains and then a highly mixed state is realized.
On the other hand, in the $\mathcal{PT}$ broken phase, the quantum trajectory analysis indicates that the quantum fluctuation is small and then the normalized magnetization is close to it for the Lindblad dynamics analysis (Fig.$\ref{tra}$ (b)). This implies that the two phases can be distinguished in terms of quantum fluctuations. Also note that this type of information can not be obtained from studying the dynamics of moments only.

To summarize, we have observed, for the open 2-spin model with the criterion of Huber \textit{et al}.. of Liouvillian $\mathcal{PT}$ symmetry, a clear change of eigenvalue structures of the Liouvillian and the dynamics in the long time limit at the $\mathcal{PT}$ symmetry breaking point. These results are similar to behaviors of NHH $\mathcal{PT}$ phase transitions and thus support  the validity of the criterion of Huber \textit{et al}. of Liouvillian $\mathcal{PT}$ symmetry.
Fig.\ref{phasedia} (b) is the phase diagram determined from the relaxation time (dynamics). Importantly, it can be determined independently of the value of physical quantities in the steady state.  Also, to our knowledge, there has been no study in which the eigenvalue structure of the Liouvillian in infinite dimension could be exactly obtained for all phases and its vivid transition could be observed. This means that our results will lead to a deeper understanding of dissipative phase transitions.   \\

%4
\section{Third quantization and $\mathcal{PT}$ symmetry for quadratic bosonic models}

\subsection{Condition of the matrix $\textbf{X}$ from criterion of Huber et al. of Liouvillian $\mathcal{PT}$ symmetry}
As shown in section I$\hspace{-.1em}$I$\hspace{-.1em}$I, the eigenvalue structure and the dynamics in the long time limit for the open 2-spin model with criterion of Huber \textit{et al}.. of Liouvillian $\mathcal{PT}$ symmetry clearly change at the $\mathcal{PT}$ symmetry breaking point. However, these results are based on exact calculations for the model and not directly derived from Liouvillian $\mathcal{PT}$ symmetry. For the Hamiltonian case, a direct relation between conventional $\mathcal{PT}$ symmetry and the dynamics (eigenvalues) has been shown $\text{\cite{MostafazadehA1}}$. In this section, for quadratic bosonic systems, we will show the criterion of Huber \textit{et al}.. of Liouvillian $\mathcal{PT}$ symmetry is rewritten into the same form as the conventional $\mathcal{PT}$ symmetry (\ref{NHHPT}) for the matrix $i\textbf{X}$.
This provides a direct relation between the criterion of Huber \textit{et al}. of Liouvillian $\mathcal{PT}$ symmetry and the dynamics of the physical quantities.

We consider a quadratic $n$ bosonic system with a linear bath whose Hamiltonian and dissipation operators are defined in Eqs.($\ref{Hamiltonian11}$), ($\ref{dissipation11}$). Importantly, when $\mathcal{P}$ in Eqs.($\ref{HuberPT3}$), ($\ref{HuberPT2}$) is a reflection operator, namely if $\mathcal{P}$ satisfies $\mathcal{P}a_i\mathcal{P}^{-1}=a_{n+1-i}$, we can write
\begin{eqnarray}
\label{Pref}
\mathcal{P}\underline{a}\mathcal{P}^{-1}=P_{n}\underline{a},
\end{eqnarray}
where
\begin{eqnarray}
P_n = \left(
    \begin{array}{cccc}
       &   & & 1 \\
       &  &1 & \\
       & \iddots &  &  \\
     1 & & & 
    \end{array}
 \right).
\end{eqnarray}

\noindent
\textbf{\textit{Theorem 3}}: 
Let us consider a quadratic bosonic system described by the Lindblad equation (\ref{lindblad}) with the Hamiltonian (\ref{Hamiltonian11}) and dissipation  operators (\ref{dissipation11}). 
Suppose that the system satisfies the criterion of Huber \textit{et al}.. of Liouvillian $\mathcal{PT}$ symmetry ($\ref{HuberPT}$) and that $\mathcal{P}$ in Eqs.($\ref{HuberPT3}$), ($\ref{HuberPT2}$)
is a reflection operator (\ref{Pref}). Then the matrix $\textbf{X}$ given by (\ref{X111}) multiplied by $i$ has conventional $\mathcal{PT}$ symmetry ($\ref{NHHPT}$), namely, $i\textbf{X}$ satisfies 
\begin{eqnarray}
\label{30}
[i\textbf{X},PT]=0
\end{eqnarray}
where the parity operator $P$ and time reversal operator $T$ are defined as
\begin{eqnarray}
\label{P}
P:=\left(
   \begin{array}{cc}
     P_{n} & 0\\
   0 & P_{n}
   \end{array}
  \right),\ \ \ \ \ \ T: i \to-i.
\end{eqnarray}
 Here $P$ means a reflection of the index of superoperators in Eq.($\rm\ref{511}$).

\begin{proof}
Since dissipation rates in Eq.($\ref{lindblad}$) are positive real numbers, the matrices $\textbf{M}$ ($\ref{M}$), $\textbf{N}$ ($\ref{N}$) and $\textbf{L}$ ($\ref{L}$) are real  unless we do not transform the elements of the vectors $\underline{l}_{\mu}$ and $\underline{k}_{\mu}$ into complex numbers as in the fermion third quantization case $\text{\cite{Prosen3}}$.

When $\mathcal{P}$ in Eqs.($\ref{HuberPT3}$), ($\ref{HuberPT2}$) is a reflection operator (\ref{Pref}),  we can rewrite Liouvillian $\mathcal{PT}$ symmetry ($\ref{HuberPT}$) as
\begin{eqnarray}
\label{HH}
\textbf{H}=P_{n}\bar{\textbf{H}}P_{n},\ \ \ \ \ \textbf{K}=P_{n}\bar{\textbf{K}}P_{n},\\
\label{NL}\textbf{N}=P_{n}\textbf{M}P_{n},\ \ \ \ \ \textbf{L}^{T}=P_{n}\textbf{L}P_{n},
\end{eqnarray}
We prove Eqs.($\ref{HH}$), ($\ref{NL}$) in Appendix D. \\
\ \  Using the relations Eqs.($\ref{HH}$), ($\ref{NL}$), we can calculate 
\begin{eqnarray}
&PT&(i\textbf{X})PT=-i(PT\textbf{X}PT)\nonumber\\
&=&\frac{-i}{2}PT\left(
   \begin{array}{cc}
     i\bar{\textbf{H}}-\textbf{N}+\textbf{M} & -2i\textbf{K}-\textbf{L}+\textbf{L}^{T}  \\
      2i\bar{\textbf{K}} -\textbf{L}+\textbf{L}^{T} & -i\textbf{H}-\textbf{N}+\textbf{M}
   \end{array}
  \right)PT\nonumber\\
  &=&\frac{-i}{2}\left(
   \begin{array}{cc}
     P_n(-i\textbf{H}-\textbf{N}+\textbf{M})P_{n} & P_n(2i\bar{\textbf{K}}-\textbf{L}+\textbf{L}^{T})P_{n}   \\
      P_n(-2i\textbf{K} -\textbf{L}+\textbf{L}^{T})P_{n}  & P_n(i\bar{\textbf{H}}-\textbf{N}+\textbf{M})P_{n} 
   \end{array}
  \right)\nonumber\\
  &=&\frac{-i}{2}\left(
   \begin{array}{cc}
     -i\bar{\textbf{H}}-\textbf{M}+\textbf{N} & 2i\textbf{K}-\textbf{L}^{T}+\textbf{L}  \\
      -2i\bar{\textbf{K}} -\textbf{L}^{T}+\textbf{L} & i\textbf{H}-\textbf{M}+\textbf{N}
   \end{array}
  \right)\nonumber\\
   &=&-i(-\textbf{X})=i\textbf{X}.
  \end{eqnarray}
\end{proof}
Note that if we had adopted $\mathbb{PT}^{\prime}(H)=H$ instead of $\mathbb{PT}(H)=H$ as the Hamiltonian $\mathcal{PT}$ symmetry ($\ref{HuberPT3}$), we would have needed $\textbf{H}$ and $\textbf{K}$ to be real to prove Theorem 1.
Since the matrix $i\textbf{X}$ has a conventional type of $\mathcal{PT}$ symmetry, if the $\mathcal{PT}$ symmetry of $i\textbf{X}$ is not broken, all the eigenvalues $\beta$ of the matrix $\textbf{X}$ ($\ref{X111}$) are pure imaginary numbers. On the other hand, if the $\mathcal{PT}$ symmetry of $i\textbf{X}$ is broken, there exist some pairs $\beta_{i},\beta_{j}$ with $\textrm{Re}[\beta_{i}]=-\textrm{Re}[\beta_{j}]\neq0$ and $\textrm{Im}[\beta_{i}]=\textrm{Im}[\beta_{j}]$. 

However, in this case, the eigenvalue structure can not be obtained from Theorem 2 because the real parts of some $\beta$ are not positive. Therefore, the physical meanings of the matrix $\textbf{X}$ ($\ref{X111}$) and $\beta$ are not obvious although we can expect that pure imaginary eigenvalues are related to a non-stationarity and negative eigenvalues are related to a divergence of physical quantities. We will investigate the time dependence of physical quantities for quadratic bosonic systems below. 

\subsection{Time evolution of quadratic bosonic systems}
In this section, we investigate the time evolution of $n$ quadratic boson systems. In quadratic bosonic systems, it is possible to calculate the higher point correlation functions from Wick's theorem if one and two point correlation functions are known. Therefore, it is sufficient to obtain the expression of the time dependence of one and two point correlation functions. We define a $2n$ vector which consists of one point correlation functions as
\begin{eqnarray}
\label{psi2}
\underline{\psi}(t):=(\textrm{Tr}[\hat{a}_{i}\rho(t)],\textrm{Tr}[\hat{a}_{i}^{\dagger}\rho(t)])^{T}
\end{eqnarray}
and a $2n \times 2n$ matrix, which consists of two point correlation functions as 
\begin{eqnarray}
\label{Z3}
\textbf{Z}(t):=\left(
   \begin{array}{cc}
    \textrm{Tr}[\hat{a}_i\hat{a}_j\rho(t)]  & \textrm{Tr}[\hat{a}_{j}^{\dagger}\hat{a}_i\rho(t)]  \\
\textrm{Tr}[\hat{a}_{i}^{\dagger}\hat{a}_j\rho(t)] & \textrm{Tr}[\hat{a}_{i}^{\dagger}\hat{a}_j^{\dagger}\rho(t)]
   \end{array}
  \right),
\end{eqnarray}
where $i,j=1,2,..,n$. 
We can derive the time derivative of the one point correlation function in Eq.($\ref{psi2}$) as 
\begin{eqnarray}
\label{psi1}
\frac{d\underline{\psi}(t)}{dt}&=&-2\textbf{X}^{T}\underline{\psi}(t).
\end{eqnarray}
This is derived in Appendix E. It can be easily solved as
\begin{eqnarray}
\label{psit}
\underline{\psi}(t)=e^{-2\textbf{X}^{T}t}\underline{\psi}(0).
\end{eqnarray}

In addition, we can also write down the time derivative of the two point correlation function in Eq.($\ref{Z3}$) as 
\begin{eqnarray}
\label{timelyapnov}
\frac{d\textbf{Z}(t)}{dt}=-2[\textbf{X}^{T}\textbf{Z}(t)+\textbf{Z}(t)\textbf{X}]+2\textbf{Y},
\end{eqnarray}
where $\textbf{Y}$ ($\rm\ref{Y}$) is a matrix which constitutes the Liouvillian $\text{\cite{Prosen4}}$.
This is also derived in Appendix E. 
In literature, the differential equation of the form Eq.($\ref{timelyapnov}$) is known as the continuous time derivative Lyapunov equation $\text{\cite{Davis,Behr}}$ and this equation can be solved as
\begin{eqnarray}
\label{Zt}
\textbf{Z}(t)=e^{-2t\textbf{X}^{T}}\textbf{Z}(0)e^{-2t\textbf{X}}+\int_{0}^{t}e^{-2s\textbf{X}^{T}}(2\textbf{Y})e^{-2s\textbf{X}}ds.\nonumber\\
\end{eqnarray}
From Eqs.($\ref{psit}$), ($\ref{Zt}$), it can be seen that the behavior of time evolution (e.g. oscillating or exponentially damping) depends only on the matrices $\textbf{X}$ and $\textbf{X}^{T}$, which have the same eigenvalues.

When the matrix $i\textbf{X}$ has conventional $\mathcal{PT}$ symmetry ($\ref{30}$), if the $\mathcal{PT}$ symmetry of $i\textbf{X}$ is not broken, all the eigenvalues of $\textbf{X}$ ($\ref{X111}$) are pure imaginary numbers and then some physical quantities oscillate after a long time. On the other hand, if the $\mathcal{PT}$ symmetry is broken, there exist some $\beta$ with a negative real part and then some physical quantities exponentially diverge. We will see simple examples in section IV. C.

We have demonstrated a direct relation between the dynamics of physical quantities and the criterion of Huber \textit{et al}. of Liouvillian $\mathcal{PT}$ symmetry for quadratic bosonic systems. 
These results also support validity of criterion of Huber et al. of Liouvillian $\mathcal{PT}$ symmetry. 

We emphasize that our arguments in this section about the $\mathcal{PT}$ symmetry ($\ref{30}$) and for the time derivative of one and two point 
correlation functions ($\ref{psi1}$), ($\ref{timelyapnov}$) can be applied to a general quadratic bosonic model with Eqs.($\ref{Hamiltonian11}$), 
($\ref{dissipation11}$). Here we note that, for 
a few restricted situations, similar results have been already discussed.  For example, in Refs.$\text{\cite{Roccati1,Purkayastha}}$ it has been found that the matrix $i\bar{\textbf{H}}-\bar{\textbf{N}}+\textbf{M}$, which is nothing but the (1,1) block matrix of $\textbf{X}$ ($\ref{X111}$), appears in the time derivative of one and two point correlation functions and also that the same matrix 
%$i\bar{\textbf{H}}-\bar{\textbf{N}}+\textbf{M}$ 
has a conventional $\mathcal{PT}$ symmetry for the case where $\textbf{K}$ in Eq.($\ref{Hamiltonian11}$) and $\textbf{L}$ ($\ref{L}$) are 0 and $n=2$ (we will investigate this case below). Also, in Refs.$\text{\cite{Arkhipov,Arkhipov1,Arkhipov3}}$ it has been found that a matrix corresponding to $i\bar{\textbf{H}}-\bar{\textbf{N}}+\textbf{M}$ appears in the time derivative of one and two point correlation functions and it has conventional anti-$\mathcal{PT}$ symmetry $\text{\cite{a}}$ for the case where $\textbf{K}=\textbf{L}=0$. Those results, however, have been obtained based on rather heuristic arguments 
and concrete calculations for specific models. In contrast, our arguments are based on the general criterion of Huber \textit{et al}. of Liouvillian $\mathcal{PT}$ symmetry or the third quantization, and hence have a much wider applicability. 

Importantly, note that the application of Theorem 3 to a quadratic boson system which appears as an HP approximation
of a spin system does not determine the Liouvillian $\mathcal{PT}$ symmetry of the original spin model. This is because the HP approximation is an approximation assuming a steady state. That is, in the $\mathcal{PT}$ broken phase, this approximation includes the information of the $\mathcal{PT}$ symmetry breaking of the steady state. Therefore, the matrix $i\textbf{X}$ in the $\mathcal{PT}$ broken phase of the open 2 spin model ($\ref{2spindefinition}$), ($\ref{2spindefinition2}$) does not satisfy the criterion of Huber \textit{et al}. of Liouvillian $\mathcal{PT}$ symmetry (and $[i\textbf{X},PT]\neq0$). However, reflecting the breaking of the steady state, $i\textbf{X}$ has conventional anti-$\mathcal{PT}$ symmetry [see e.g. Eq.($\rm{\ref{XX}}$)].
Also, the HP approximation can not be applied to the $\mathcal{PT}$ phase since it is the disordered phase. Therefore, the $\mathcal{PT}$ symmetry of $i\textbf{X}$ should not be adopted as the definition of $\mathcal{PT}$ symmetry of the large spin systems.

%Therefore, the PT symmetry of $i\textbf{X}$ should not be adopted as the definition of PT symmetry of the large spin systems as in Ref.$\text{\cite{QED}}$, although the open 2 spin model can be mapped to the quadratic bosonic model with the HP approximation.

%However, physically considering, the system (i.e. its corresponding Liouvillian) should satisfy the PT symmetry in the region where the dissipation is balanced regardless of whether the PT symmetry of the state is broken or not. 

\subsection{Example 1: An open 2-boson model with balanced gain and loss}
We consider an open 2-component quadratic bosonic model with the criterion of Huber \textit{et al}. of Liouvillian $\mathcal{PT}$ symmetry ($\ref{HuberPT}$) as
\begin{eqnarray}
\label{bosonn}
H&=&g(c_A^{\dagger}c_B+\rm{H.c.}),\\
\dot{\rho}=\hat{\mathcal{L}}\rho&=&-i[H,\rho]+\Gamma\mathcal{D}[c_A]\rho+\Gamma\mathcal{D}[c_B^{\dagger}]\rho.
\end{eqnarray}
The Hamiltonian and types of dissipations are the same as the HP approximated 2-spin model in the FM $\ket{\Uparrow\Uparrow}$ phase in section I$\hspace{-.1em}$I$\hspace{-.1em}$I which has been studied in detail in Refs.$\text{\cite{Dast1,Roccati1}}$.%In particular, in Ref.$\text{\cite{Roccati1}}$, it has been found that the PT symmetric matrices appear when writing down the time derivative of the one and two point correlation functions (these facts correspond to that $i\textbf{X}$ and $i\textbf{X}^{T}$ have conventional PT symmetry). However, those results have not been derived from the viewpoint of criterion of Huber et al. of Liouvillian PT symmetry or the third quantization and a general quadratic bosonic model in Eqs.($\ref{Hamiltonian}$), ($\ref{dissipation}$) has not been investigated. \\

Calculating one point and two point correlation functions from Eqs.($\ref{psit}$), ($\ref{Zt}$), it can be seen that the dynamics changes at a point $\Gamma=g$. (We give the expression of the time dependence of one and two correlation functions for this model in Appendix E.) Fig.$\ref{fig2boson}$ (a), (b) show the time evolution of one point correlation function $\braket{c_{A}}$ above and below the transition point $\Gamma=g$. In the $\mathcal{PT}$ unbroken phase (of the matrix $i\textbf{X}$), the one point correlation function oscillates in the long time limit, while in the $\mathcal{PT}$ broken phase, it exponentially diverges.

\begin{figure}[ht]
  \begin{center}
   \vspace*{1.6cm}
     \hspace*{-6.2cm}
\includegraphics[bb=0mm 0mm 90mm 150mm,width=0.36\linewidth]{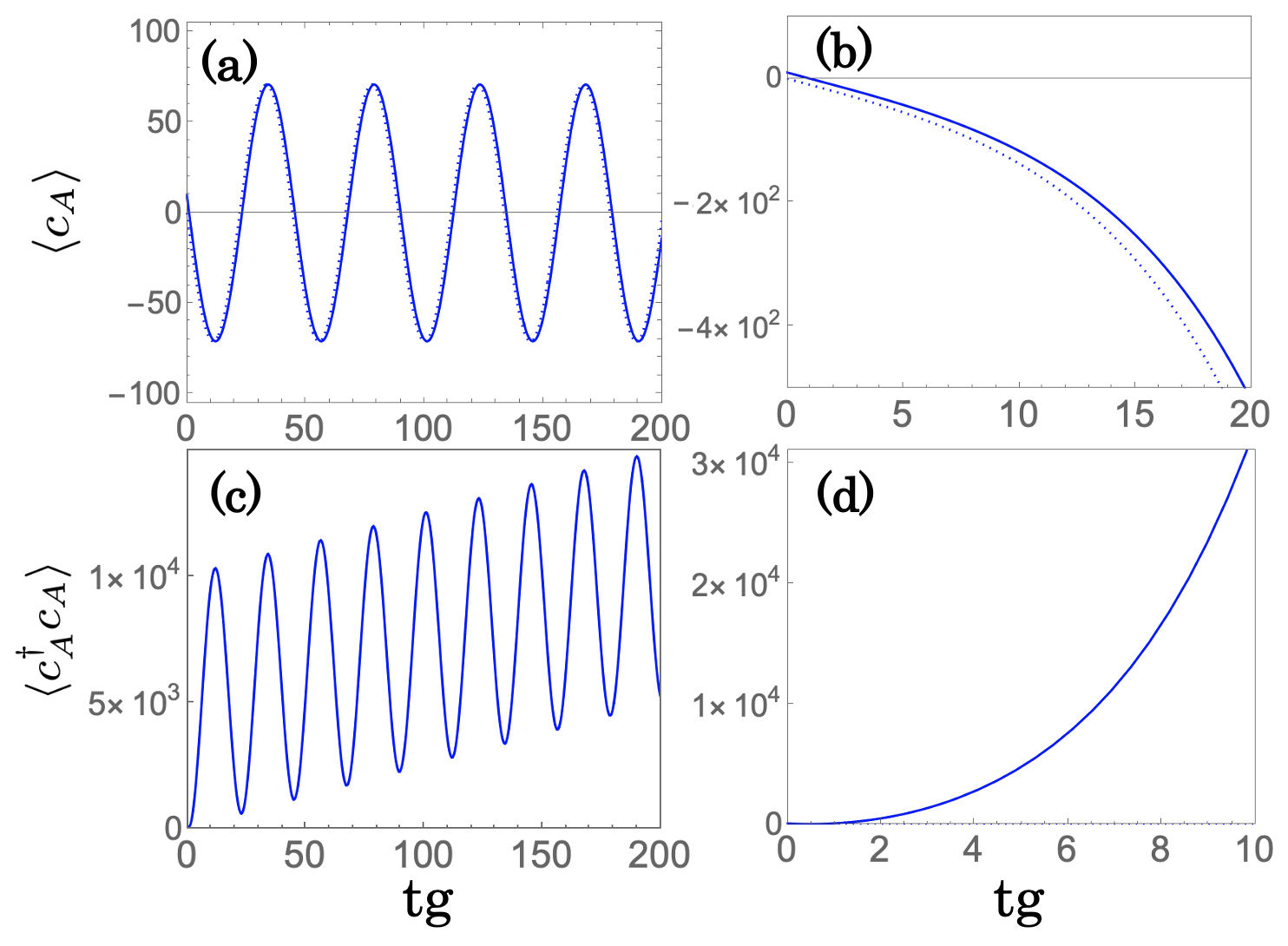}
    \caption{Time evolution of the one and two point correlation function; (a),(b) $\braket{c_{A}}$ (c),(d) $\braket{c_{A}^{\dagger}c_{A}}$. We set parameters as (a),(c) $\Gamma=0.99, g=1$ in the $\mathcal{PT}$ phase (b),(d) $\Gamma=1.01, g=1$ in the $\mathcal{PT}$ broken phase, and $\braket{c_{A}}(0)=\braket{c_{B}}(0)=10$, $\braket{c_{A}^{\dagger}c_{A}}(0)=\braket{c_{B}^{\dagger}c_{B}}(0)=100$, $\braket{c_{A}^{\dagger}c_{B}}(0)=\braket{c_{B}^{\dagger}c_{A}}(0)=0$. Solid and doted lines indicate real and imaginary parts, respectively.}
    \label{fig2boson}
  \end{center}
\end{figure}

Fig.$\ref{fig2boson}$ (c), (d) show the time evolutions of the two point correlation function $\braket{c_{A}^{\dagger}c_{A}}$ above and below the transition point $\Gamma=g$. In the $\mathcal{PT}$ unbroken phase, the two point correlation function diverges linearly with oscillation, while in the $\mathcal{PT}$ broken phase, it diverges exponentially without oscillation. Note that these behaviors are different from those of the open 2-spin model in section I$\hspace{-.1em}$I$\hspace{-.1em}$I because bosonic models are upper unbounded but spin models are upper bounded. \\

\subsection{Example 2: An open 2-mode Rabi model with balanced gain and loss}
The quantum Rabi model is a simple model including the coupling between two-level atom and quantized cavity mode $\text{\cite{Braak,Hwang2,Puebla,Zhang,Cai}}$, and it has been widely investigated even in dissipative systems $\text{\cite{Hwang,Malekakhlagh,Wang}}$. %Also, Rabi model with two cavity modes has been also studied$\text{\cite{}}$.
Here, we review an open 2-mode Rabi model with balanced gain and loss studied in Ref.$\text{\cite{Wang}}$. Hamiltonian and Lindblad equation of this model are defined as 
\begin{eqnarray}
\label{RabiH}
H&=&\omega(c_A^{\dagger}c_A+c_B^{\dagger}c_B)+\frac{\Omega}{2}\sigma_{z}\nonumber\\
&-&\lambda(c_A^{\dagger}+c_A)\sigma_{x}-\lambda(c_B^{\dagger}+c_B)\sigma_{x},\\
\dot{\rho}=\hat{\mathcal{L}}\rho&=&-i[H,\rho]+\Gamma\mathcal{D}[c_A]\rho+\Gamma\mathcal{D}[c_B^{\dagger}]\rho,
\end{eqnarray}
where $\sigma_{x}$, $\sigma_{z}$ are Pauli matrices for a two level system. $\lambda$ is the coupling strength between the cavity and the atom, and $\Omega$ is the qubit transition frequency. We introduce
a frequency ratio $\eta:=\Omega/\omega$ and a dimensionless coupling constant $g:=2\lambda/\sqrt{\omega\Omega}$. Hereafter, we consider the limit of ultrastrong coupling, $\lambda/\omega\gg1$,
and extremely large detuning, $\eta\gg1$, but keep the coupling
constant $g\sim \mathcal{O}(1)$ finite.
When $g$ increases, a transition from the normal phase to the superradiant phase occurs at $g=\sqrt{(1+\frac{\Gamma^{2}}{\omega^{2}})/2}=:g_{c}$ $\text{\cite{Wang}}$. In this paper, we consider only the normal phase ($g<g_{c}$).

Applying a unitary transformation
\begin{eqnarray}
U=\exp[ig\eta^{-\frac{1}{2}}(c_{A}+c_{A}^{\dagger}+c_{B}+c_{B}^{\dagger})\sigma_{y}]
\end{eqnarray}
to the Hamiltonian ($\ref{RabiH}$), it can be transformed to
\begin{eqnarray}
U^{\dagger}HU&=&\omega(c_A^{\dagger}c_A+c_B^{\dagger}c_B)+\frac{\Omega}{2}\sigma_{z}\nonumber\\
&-&\frac{\omega g^{2}}{4}(c_A+c_A^{\dagger}+c_B+c_B^{\dagger})^2\sigma_{z}+\mathcal{O}(\eta^{-\frac{1}{2}}).
\end{eqnarray}
Here, the unitary transformation does not affect the form of Lindblad operators. Then, tracing out the qubit degree of freedom on the ground state of the qubit $\ket{\Downarrow}$, the Lindblad equation is transformed as 
\begin{eqnarray}
\label{RL}
\dot{\rho}_{np}=-i[H_{np},\rho_{np}]+\Gamma\mathcal{D}[c_A]\rho_{np}+\Gamma\mathcal{D}[c_B^{\dagger}]\rho_{np},
\end{eqnarray}
with
\begin{eqnarray}
\label{RL2}
H_{np}&=&\bra{\Downarrow}U^{\dagger}HU\ket{\Downarrow}\nonumber\\
&=&\omega(c_A^{\dagger}c_A+c_B^{\dagger}c_B)-\frac{\omega g^{2}}{4}(c_A+c_A^{\dagger}+c_B+c_B^{\dagger})^2,\nonumber\\
\end{eqnarray}
where $\rho_{np}=\bra{\Downarrow}U^{\dagger}\rho U\ket{\Downarrow}$. This Lindblad equation ($\ref{RL}$) and Hamiltonian ($\ref{RL2}$) satisfy the criterion of Huber \textit{et al}. of Liouvillian $\mathcal{PT}$ symmetry ($\ref{HuberPT}$). Also, since this model is transformed to one which has the quadratic Hamiltonian with linear bath operators, we can apply the third quantization. The matrix $\textbf{X}$ ($\ref{X111}$) for this model can be calculated as 
\begin{eqnarray}
\label{X2}
\textbf{X}=
\frac{1}{2}\left(
   \begin{array}{cccc}
     i\omega^{\prime}+\Gamma & -ir&ir&ir  \\
     -ir& i\omega^{\prime}-\Gamma &ir&ir  \\
    -ir&-ir& -i\omega^{\prime}+\Gamma &ir  \\
    -ir& -ir&ir& -i\omega^{\prime}-\Gamma  
   \end{array}
  \right),\nonumber\\
\end{eqnarray}
where $r:=\omega g^{2}/2$, $\omega^{\prime}:=\omega-r$. Importantly, the matrix ($\ref{X2}$) satisfies Eq.($\ref{30}$) and $\textbf{K}\neq0$. Then, the eigenvalues $\beta$ are calculated as
\begin{eqnarray}
\beta=\pm i \sqrt{A\pm2\sqrt{B}},
\end{eqnarray}
where
\begin{eqnarray}
A&=&\omega^{2}-\omega^{2}g^{2}-\Gamma^{2},\\
B&=&\omega^{2}(\frac{\omega^{2}g^{4}}{4}+g^{2}\Gamma^{2}-\Gamma^{2}).
\end{eqnarray}
Here, $A\pm2\sqrt{B}=0$ is equivalent to $g=g_{c}$, and $B=0$ is equivalent to
\begin{eqnarray}
g^{2}=2\left(\sqrt{\frac{\Gamma^{4}}{\omega^{4}}+\frac{\Gamma^{2}}{\omega^{2}}}-\frac{\Gamma^{2}}{\omega^{2}}\right).
\end{eqnarray}
When $B>0$ and $A\pm2\sqrt{B}>0$, all $\beta$ are pure imaginary numbers, namely the state oscillates in the long time limit. On the other hand, when $B<0$, all $\beta$ are not pure imaginary numbers. Also, since we consider the normal phase, the state decays to the steady state with oscillation. This transition is caused by the $\mathcal{PT}$ symmetry breaking of the matrix $\textbf{X}$ ($\ref{X2}$). Similarly, the $\mathcal{PT}$ symmetry breaking occurs in the superradiant phase.

%\begin{eqnarray}
%g^{2}=2\left(\sqrt{\frac{\Gamma^{4}}{\omega^{4}}+\frac{\Gamma^{2}}{\omega^{2}}}-\frac{\Gamma^{2}}{\omega^{2}}\right)=:g_{c}^{\prime\  2}
%\end{eqnarray}
%$g_{c}>g_{c}^{\prime}$

%6

\section{Summary and Discussion}
We investigated $\mathcal{PT}$ phase transitions in open quantum systems with the third quantization, in particular, eigenvalue structure of the Liouvillian and time dependence of physical quantities for open spin and bosonic systems. First, we applied it to the open 2-spin model with the criterion of Huber \textit{et al}. of Liouvillian $\mathcal{PT}$ symmetry to investigate the eigenvalue structure and dynamics in the long time limit below and above the $\mathcal{PT}$ symmetry breaking point. In particular, we found that in the $\mathcal{PT}$ unbroken phase, some eigenvalues are pure imaginary numbers while in the $\mathcal{PT}$ broken phase, all eigenvalues are real. From this result, it could be analytically shown for open systems with Lindblad dynamics that the evolution of physical quantities changes from an oscillatory to an overdamped behavior at the $\mathcal{PT}$ symmetry breaking point. These results support the validity of criterion of Huber \textit{et al}. of Liouvillian $\mathcal{PT}$ symmetry.

Next, we showed that the matrix $i\textbf{X}$ has conventional $\mathcal{PT}$ symmetry if the Liouvillian satisfies the criterion of Huber \textit{et al}. of Liouvillian $\mathcal{PT}$ symmetry for quadratic bosonic models. We remark that this result shows that the definition of Huber \textit{et al}. is rewritten in the same form as the conventional condition of $\mathcal{PT}$ symmetry (\ref{NHHPT}).
% although it has been phenomenological condition so far. 
Furthermore, we derived the time dependence of one and two point correlation functions for bosonic systems and showed that the dynamics of physical quantities depend on only the eigenvalues of the matrix $\textbf{X}$. From these results, we clarified the relation between the criterion of Huber \textit{et al}. of Liouvillian $\mathcal{PT}$ symmetry and physical quantities. These results also support the validity of the criterion of Huber \textit{et al}. of Liouvillian $\mathcal{PT}$ symmetry.

We also found that the $\mathcal{PT}$ symmetry breaking point for the open 2-spin model is a Liouvillian exceptional point if it is considered as a limit from the FM phases and that 
the Liouvillian gap is closed at the phase boundary. % (See Appendices A and C).
However, it has not been completely clarified whether it is actually an exceptional point or a degenerate point. For example, in the limit from the AFM phase, we did not show that the $\mathcal{PT}$ symmetry breaking point is a Liouvillian exceptional point. For the $\mathcal{PT}$ symmetric Hamiltonian case, $\mathcal{PT}$ phase transition points are known to be exceptional points, and many unconventional phenomena and applications $\text{\cite{Makris,Chen,Guo,Ramezani}}$ in the vicinity of $\mathcal{PT}$ phase transition points occur from the properties of the exceptional points. Therefore, it is very important to investigate this question and we plan to do so in near future. 
Note that the studies of Liouvillian exceptional points have been actively conducted recently $\text{\cite{Minganti1,MingantiH,Arkhipov,Arkhipov1,Arkhipov3,Jan}}$. 

Also, as a natural extension, we may consider application of the third quantization to an open $2N$-spin model which has alternately pumping spins in opposite directions. The phase diagram and physical quantities of the open $2N$-spin model have been investigated in Ref.$\text{\cite{Huber1}}$. Furthermore, the topological properties for the Liouvillian have also been studied recently $\text{\cite{He,van,Dangel,Lieu1}}$, so it may be possible to analyze them in the open $2N$-spin system as well. Also, the application to fermions is really interesting. However, unlike large spin systems and bosonic systems, in the fermionic case a matrix to determine full Liouvillian eigenvalues does not generally have $\mathcal{PT}$ symmetry or anti-$\mathcal{PT}$ symmetry. The relation between fermionic third quantization $\text{\cite{Prosen3}}$ and Liouvillian $\mathcal{PT}$ symmetry is left for future work.

\begin{acknowledgments}
We thank Kohei Yamanaka for fruitful discussions. YN also acknowledges financial support from Advanced Research Center for Quantum Physics and Nanoscience and Tokyo Tech Academy for Convergence of Materials and Informatics.
The work of TS was supported by JSPS KAKENHI Grants No. JP16H06338, No. JP18H01141, No. JP18H03672, No. JP19L03665. 
\end{acknowledgments}

\onecolumngrid
\appendix
\newpage

%A.1
\section{Third quantization}
\label{A1}
%We review the framework of the third quantization which is a general method to solve the Lindblad equation for open quadratic systems with linear bath operators $\text{\cite{Prosen4,Prosen3}}$.
In the main text, we reviewed the case where the matrix $\textbf{X}$ ($\rm\ref{X111}$) is diagonalizable. Here we consider the case where the matrix $\textbf{X}$ is nondiagonalizable in analogy to the fermionic case $\text{\cite{Prosen11}}$.
First, we define two vector spaces $\mathcal{K}$ and $\mathcal{K}^{\prime}$, where $\mathcal{K}$ contains trace class operators (e.g. density matrices) and $\mathcal{K}^{\prime}$ contains unbounded operators that we need as physical observables. Hereafter, we adopt Dirac notation and write an element of $\mathcal{K}$ as $\ket{\rho}$ and an element of $\mathcal{K}^{\prime}$ as $(A|$, where
their inner product gives the expectation value of an observable $A$ for a state $\rho$,
\begin{equation}
\label{inner}
(A\ket{\rho}=\textrm{tr}A\rho.
\end{equation}

Hamiltonian and dissipation operators for an arbitrary quadratic system of $n$ bosons
with linear bath operators can be written in Eqs.($\ref{Hamiltonian11}$), ($\ref{dissipation11}$) as
\begin{eqnarray}
\label{Hamiltonian1}
H&=&\underline{a}^{\dagger}\cdot\textbf{H}\underline{a}+\underline{a}\cdot\textbf{K}\underline{a}+\underline{a}^{\dagger}\cdot\bar{\textbf{K}}\underline{a}^{\dagger},\\
\label{dissipation1}L_{\mu}&=&\underline{l}_{\mu}\cdot\underline{a}+\underline{k}_{\mu}\cdot\underline{a}^{\dagger}.
\end{eqnarray}
Then, the Liouvillian consists of the two matrices \textbf{X} ($\ref{X111}$) and \textbf{Y} ($\ref{Y111}$)
\begin{eqnarray}
\label{X1}
\textbf{X}=\frac{1}{2}\left(
   \begin{array}{cc}
     i\bar{\textbf{H}}-\bar{\textbf{N}}+\textbf{M} & -2i\textbf{K}-\textbf{L}+\textbf{L}^{T}  \\
      2i\bar{\textbf{K}} -\bar{\textbf{L}}+\bar{\textbf{L}}^{T} & -i\textbf{H}-\textbf{N}+\bar{\textbf{M}}
   \end{array}
  \right),
\end{eqnarray}
and
\begin{eqnarray}
\label{Y}
\textbf{Y}=\frac{1}{2}\left(
   \begin{array}{cc}
     -2i\bar{\textbf{K}}-\bar{\textbf{L}}-\bar{\textbf{L}}^{T}  & 2\textbf{N} \\
      2\textbf{N}^{T} & 2i\textbf{K}-\textbf{L}-\textbf{L}^{T} 
   \end{array}
  \right).
\end{eqnarray}
Here the matrices $\textbf{M}$, $\textbf{N}$ and $\textbf{L}$ are defined in Eqs.($\ref{M}$),($\ref{N}$) and ($\ref{L}$) as
\begin{eqnarray}
\label{M1}\textbf{M}&:=&\sum_{\mu}\underline{l}_{\mu}\otimes\underline{\bar{l}}_{\mu}=\textbf{M}^{\dagger},\\ 
\label{N1}\textbf{N}&:=&\sum_{\mu}\underline{k}_{\mu}\otimes\underline{\bar{k}}_{\mu}=\textbf{N}^{\dagger},\\ 
\label{L1}\textbf{L}&:=&\sum_{\mu}\underline{l}_{\mu}\otimes\underline{\bar{k}}_{\mu}.
\end{eqnarray}

For the nondiagonalizable case, $\textbf{X}$ ($\rm\ref{X1}$) can be written in analogy to the fermionic case $\text{\cite{Prosen11}}$ as
\begin{eqnarray}
\label{202}
\textbf{X}=\textbf{P}{\boldsymbol{\Delta}}\textbf{P}^{-1},
\end{eqnarray}
where $\textbf{P}$ is a $2n$ $\times$ $2n$ matrix and ${\boldsymbol{\Delta}}=\oplus_{j=1}^{q}{\boldsymbol{\Delta}}_{l_{j}}( \beta_{j})$ is a direct sum of Jordan blocks
\begin{eqnarray}
{\boldsymbol{\Delta}}_{l}( \beta):=\left(
   \begin{array}{ccccc}
     \beta  & 1& & & \\
       & \beta&\ddots& & \\
       & & \ddots& &1 \\
    & &   & &\beta 
   \end{array}
  \right),
\end{eqnarray}
for the direct eigenvalues $\beta_{j}$, and $q$ is the number of blocks and $l_{j}$ is the block size which satisfy
\begin{eqnarray}
\sum_{j=1}^{q}l_{j}=2n.
\end{eqnarray}
In this case, we can show that the Liouvillian ($\rm\ref{lio1}$) can be also written as 
\begin{eqnarray}
\label{EP}
\hat{\mathcal{L}}=-2\sum_{j=1}^{q}\left[\beta_{j}\sum_{r=1}^{l_{j}}\hat{\zeta}^{\prime}_{j,r}\hat{\zeta}_{j,r}+\sum_{r=1}^{l_{j}-1}\hat{\zeta}^{\prime}_{j,r+1}\hat{\zeta}_{j,r}\right],
\end{eqnarray}
where normal master-modes $\hat{\underline{\zeta}}=(\zeta_{1,1}, \cdots , \zeta_{1,l_{1}}, \zeta_{2,1}, \cdots , \zeta_{q,l_{q}})$ and $\hat{\underline{\zeta}}^{\prime}=(\zeta_{1,1}^{\prime}, \cdots , \zeta_{1,l_{1}}^{\prime}, \zeta_{2,1}^{\prime}, \cdots , \zeta_{q,l_{q}}^{\prime})$ are defined by ($\rm\ref{241}$) with
\begin{eqnarray}
\label{zeta1}
[\hat{\zeta}_{j,r},\hat{\zeta}^{\prime}_{k,s}]=\delta_{j,k}\delta_{r,s},\ \ \ \ \ [\hat{\zeta}_{j,r},\hat{\zeta}_{k,s}]=[\hat{\zeta}_{j,r}^{\prime},\hat{\zeta}^{\prime}_{k,s}]=0.
\end{eqnarray}
Here, $j,k=1,2,...,m$ and $r,s=1,2,...,l_{j}$.

\textbf{\textit{Theorem A1}}: If the real parts of all the eigenvalues are positive, $\forall j,\rm{Re} [\beta_{j}]>0$, the full spectrum of the Liouvillian is given in the form, 
\begin{eqnarray}
\label{ne}
\lambda_{\underline{m}}=-2\sum_{j=1}^{q}\sum_{r=1}^{l_{j}}m_{j,r}\beta_{j},
\end{eqnarray}
in terms of a $2n$ component multi-index of super-quantum numbers $\underline{m}=(m_{1,1},\ \cdots\,  m_{1,l_{1}},\ m_{2,1},\ \cdots\, m_{q,l_{q}})\in\mathbb{Z}_{+}^{2n}$. 
\begin{proof}
We write the Liouvillian ($\rm\ref{EP}$) as $\hat{\mathcal{L}}=\hat{\mathcal{L}}_{0}+\hat{\mathcal{M}}$, where $\hat{\mathcal{L}}_{0}:=-2\sum_{j=1}^{q}\beta_{j}\hat{\mathcal{N}}_{j}$, $\hat{\mathcal{M}}:=-2\sum_{j=1}^{q}\hat{\mathcal{M}}_{j}$ and $\hat{\mathcal{N}}_{j}:=\sum_{r=1}^{l_{j}}\hat{\zeta}^{\prime}_{j,r}\hat{\zeta}_{j,r}$, $\hat{\mathcal{M}}_{j}:=\sum_{r=1}^{l_{j}-1}\hat{\zeta}^{\prime}_{j,r+1}\hat{\zeta}_{j,r}$. 
Since $\hat{\mathcal{M}}_{j}$ is clearly nilpotent and $[\hat{\mathcal{M}}_{j},\hat{\mathcal{M}}_{k}]=0$, $\hat{\mathcal{M}}$ is also nilpotent. Moreover, $[\hat{\mathcal{L}}_{0},\hat{\mathcal{M}}]=0$ because all the terms 
in the definitions of $\hat{\mathcal{L}}_{0}$ and $\hat{\mathcal{M}}$ above commute $[\hat{\mathcal{N}}_{j},\hat{\mathcal{M}}_{k}]=0$, so $\hat{\mathcal{L}}$ and $\hat{\mathcal{L}}_{0}$ have the same spectra. The eigenvalues of $\hat{\mathcal{L}}_{0}$ are given in ($\rm\ref{newbasis}$), so we  get (\ref{ne}).
\end{proof}
Moreover, we can say that an exceptional point of $\textbf{X}$ is also a Liouvillian exceptional point. In fact, at an exceptional point of $\textbf{X}$ with order $p$ (where $p$ eigenvectors coalesce), we can find that some terms included in one and two point correlation functions are proportional to $t^{p-1}$ from the expression of the time dependence of one and two point correlation functions ($\ref{psit}$), ($\ref{Zt}$). Furthermore, some terms in $2n$ and $2n-1$ point correlation functions are also proportional to $t^{n(p-1)}$. This implies that this point is also a Liouvillian exceptional point with infinite order since the time dependence of the density operator (physical quantities) is exponential as Eq.($\ref{110}$) if the Liouvillian is diagonalizable. Similar discussions have been conducted and it has been shown that there are infinite order Liouvillian exceptional points for other quadratic bosonic models in Refs.$\text{\cite{Arkhipov,Arkhipov1,Arkhipov3}}$. 
%\sout{Note, however, that it is very complicated to ''actually'' perform the calculation of high order point correlation functions.}

Also, it is known that at an exceptional point, the eigenspace cannot be spanned only by the eigenstates, but generalized eigenstates are required $\text{\cite{Kanki}}$. If $\lambda_{i}$ is an eigenvalue which $n+1$ eigenmodes coalesce, an eigenmode $\rho_{i}$ and generalized eigenstates $\rho_{i}^{(1)}$, $\rho_{i}^{(2)}$, ..., $\rho_{i}^{(n)}$ satisfy
\begin{eqnarray}
\label{EPP}
(\hat{\mathcal{L}}-\lambda_{i})\rho_{i}=0,\ \ \ \ \ \ \ \ 
(\hat{\mathcal{L}}-\lambda_{i})\rho^{(1)}_{i}=\rho_{i},\ \ \ \ \cdots, \ \ 
(\hat{\mathcal{L}}-\lambda_{i})\rho^{(n)}_{i}=\rho_{i}^{(n-1)}.
\end{eqnarray}
For the case of Eq.($\rm\ref{EP}$), it has not been known how to get all the eigenstates and all the generalized eigenstates in general. 
Here we just see what happens for a simple example of  $q=1$, $l_{1}=2$, namely when there is only one Jordan block of size two. In this case the  Liouvillian (\ref{EP}) is written as 
\begin{eqnarray}
\hat{\mathcal{L}}=-2\left[\beta_{1}\sum_{r=1}^{2}\hat{\zeta}^{\prime}_{1,r}\hat{\zeta}_{1,r}+\hat{\zeta}^{\prime}_{1,2}\hat{\zeta}_{1,1}\right]
\end{eqnarray}
with the eigenvalue (\ref{ne}) given by $\lambda_{\underline{m}}=-2(m_{1,1}+m_{1,2})\beta_{1}$. 
Moreover all the eigenmodes $\rho_{(0,m_{1,2})}$ and all the generalized eigenmodes $\rho_{(m_{1,1},m_{1,2})}^{\prime}, m_{1,1}\neq0$ are found explicitly as 
\begin{eqnarray}
\rho_{(0,m_{1,2})}=(\hat{\zeta}^{\prime}_{1,2})^{m_{1,2}}\rho_{ss}
\end{eqnarray}
and
\begin{eqnarray}
\rho_{(m_{1,1},m_{1,2})}^{\prime}=\frac{1}{(-2)^{m_{1,1}}{m_{1,1}}!}(\hat{\zeta}^{\prime}_{1,2})^{m_{1,2}}(\hat{\zeta}^{\prime}_{1,1})^{m_{1,1}}\rho_{ss},
\quad m_{1,1}\neq0
\end{eqnarray}
respectively. In fact, we can directly check that all  $\rho_{(0,m_{1,2})}$ and $\rho_{(m_{1,1},m_{1,2})}^{\prime}, m_{1,1}\neq0$ above satisfy the relations 
($\rm\ref{EPP}$) for this particular case as
\begin{eqnarray}
[\hat{\mathcal{L}}+2m_{1,2}\beta_{1}]\rho_{(0,m_{1,2})}=[\hat{\mathcal{L}}+2m_{1,2}\beta_{1}](\hat{\zeta}^{\prime}_{1,2})^{m_{1,2}}\rho_{ss}=0
\end{eqnarray}
and
\begin{eqnarray}
[\hat{\mathcal{L}}+2(m_{1,1}+m_{1,2})\beta_{1}]\rho_{(m_{1,1},m_{1,2})}^{\prime}&=&\frac{1}{(-2)^{m_{1,1}-1}{(m_{1,1}-1)}!}(\hat{\zeta}^{\prime}_{1,2})^{m_{1,2}+1}(\hat{\zeta}^{\prime}_{1,1})^{(m_{1,1}-1)}\rho_{ss}\nonumber\\
&=&\rho_{(m_{1,1}-1,m_{1,2}+1)}^{\prime}.
\end{eqnarray}
Hence associated with an eigenvalue of the form $-2n\beta_1,n\in\mathbb{Z}_+$, are one eigenmode $\rho_{(0,n)}$ and $n$ generalized eigenmodes $\rho_{(m_{1,1},m_{1,2})}^{\prime},\  m_{1,1}+m_{1,2}=n,\ m_{1,1}\in\mathbb{Z}_+,\ m_{1,2}\in\mathbb{N}$ as a result of coalescence of $n+1$ eigenmodes.
Therefore, from the full Liouvillian spectrum, we can easily and rigorously find that there are eigenvalues at which an infinite number of eigenmodes coalesce since $n$ can be arbitrarily large. Note that there do not appear high-order effects on one and two point correlation functions even at a Liouvillian exceptional point with infinite order in this case since the time dependence of one and two point correlation functions ($\ref{psit}$), ($\ref{Zt}$) is determined only by $\textbf{X}$ ($\rm\ref{X1}$) and $\textbf{X}^T$. 

The order of exceptional points is crucially important for certain realistic technologies which apply concepts and results to $\mathcal{PT}$ phase transitions. For example, it is known that the sensitivity of a $\mathcal{PT}$ symmetric high-sensitivity sensor is determined by the order of exceptional points (in the classical regimes) $\text{\cite{Chen}}$, and the enhancement of sensitivity occurs for any models which have exceptional points. Therefore, it is also important to analyze the order of exceptional points even for quadratic systems when considering the those realizations in the quantum regime, where quantum noise such as a quantum jump must be taken into account. In fact, in several references $\text{\cite{Minganti1,Jan}}$, the sensitivity of perturbations at Liouvillian exceptional points has been discussed. From this point of view, the derivation of the full Liouvillian spectrum (like in our work) is also important.

\section{Holstein-Primakoff approximation}
\label{A2}
The Holstein-Primakoff (HP) transformation $\text{\cite{Holstein}}$ provides an exact mapping of the spin operators to bosonic operators. First, we consider the case where the state $\ket{-S}$ transforms to a bosonic vacuum state with the total spin $S$. Then, the HP transformation is defined as 
\begin{eqnarray}
\label{SSS}
S^{+}=c^{\dagger}\sqrt{2S-c^{\dagger}c},\ \ \ \ \ \ S^{-}=\sqrt{2S-c^{\dagger}c}c,\ \ \ \ \ \ S^{z}=-S+c^{\dagger}c,
\end{eqnarray}
where $c$ ($\textit{resp}$. $c^{\dagger}$) is the bosonic annihilation ($\textit{resp}$. creation) operator. Also, for $\braket{c^{\dagger}c}/(2S)\ll 1$, we can approximate the spin operators in Eq.($\rm\ref{SSS}$) by
\begin{eqnarray}
\label{-}
S^{+}\simeq \sqrt{2S}c^{\dagger},\ \ \ \ \ \ S^{-}\simeq\sqrt{2S}c,\ \ \ \ \ \ S^{z}=-S+c^{\dagger}c.
\end{eqnarray}
Similarly, we consider the case where the state $\ket{S}$ transforms to a bosonic vacuum state and $\braket{c^{\dagger}c}/(2S)\ll 1$, and then we can approximate the spin operators by
\begin{eqnarray}
\label{+}
S^{+}\simeq \sqrt{2S}c,\ \ \ \ \ \ S^{-}\simeq\sqrt{2S}c^{\dagger},\ \ \ \ \ \ S^{z}=S-c^{\dagger}c.
\end{eqnarray}

%A.2
\section{Calculation of eigenvalue structure for the open 2-spin model with the third quantization}
\label{A3}

\subsection{Third quantization for the AFM phase}
First, we apply the third quantization to the AFM phase. The Hamiltonian and Lindblad equation can be written as 
\begin{eqnarray}
\label{h11}
H&=&g(c_Ac_B+\rm{H.c.}),\\
\label{lind1}\dot{\rho}=\hat{\mathcal{L}}\rho&=&-i[H,\rho]+\Gamma_g\mathcal{D}[c_A]\rho+\Gamma_l\mathcal{D}[c_B]\rho
\end{eqnarray}
with the HP approximation ($\rm\ref{-}$), ($\rm\ref{+}$) for $S\gg1$. Here, keep in mind that the HP approximated model ($\rm\ref{h11}$), ($\rm\ref{lind1}$) doesn't satisfy the definition of Liouvillian $\mathcal{PT}$ symmetry ($\rm\ref{HuberPT}$). From Lindblad equation ($\rm\ref{lind1}$), we can write the dissipation operators ($\rm\ref{dissipation1}$) as
\begin{eqnarray}
L_{1}=\sqrt{\Gamma_{g}}c_{A}=\left(
   \begin{array}{c}
      \sqrt{\Gamma_{g}}  \\
      0
    \end{array}
  \right)\cdot\left(
   \begin{array}{c}
     c_{A}  \\
      c_{B}
    \end{array}
  \right)=\underline{l_{1}}\cdot\underline{c},\\
  L_{2}=\sqrt{\Gamma_{l}}c_{B}=\left(
   \begin{array}{c}
     0  \\
       \sqrt{\Gamma_{l}}
    \end{array}
  \right)\cdot\left(
   \begin{array}{c}
     c_{A}  \\
      c_{B}
    \end{array}
  \right)=\underline{l_{2}}\cdot\underline{c}.
\end{eqnarray}
Therefore, we can calculate the matrices $\textbf{K}$ in Eq.($\rm\ref{Hamiltonian1}$) and $\textbf{M}$ ($\rm\ref{M1}$) as
\begin{eqnarray}
\textbf{K}=\frac{g}{2}\left(
   \begin{array}{rr}
      0 & 1  \\
      1 & 0 
    \end{array}
  \right),\ \ \ \ \ \textbf{M}=\left(
   \begin{array}{rr}
      \Gamma_{g} & 0  \\
      0 & \Gamma_{l}
    \end{array}
  \right)
\end{eqnarray}
and $\textbf{H}=\textbf{N}=\textbf{L}=\textbf{0}$. Therefore, we find that the matrices $\textbf{X}$ ($\rm\ref{X1}$) and $\textbf{Y}$ ($\rm\ref{Y}$) are written as
\begin{eqnarray}
\label{XX}
\textbf{X}=\frac{1}{2}\left(
    \begin{array}{rrrr}
     \Gamma_g & 0&0&-ig\\
      0& \Gamma_l &-ig&0\\
   0 & ig&\Gamma_g&0\\
      ig &0&0& \Gamma_l 
    \end{array}
  \right)
\end{eqnarray}
and
\begin{eqnarray}
\label{AMY}
\textbf{Y}=\frac{1}{2}\left(
    \begin{array}{rrrr}
     0& -ig&0&0\\
      -ig& 0 &0&0\\
   0 & 0&0&ig\\
      0 &0&ig& 0 
    \end{array}
  \right),
\end{eqnarray}
respectively. \textit{Here, keep in mind that the matrix i$\textbf{X}$ doesn't have $\mathcal{PT}$ symmetry but anti-$\mathcal{PT}$ symmetry}. Note that the matrix $\textbf{X}$ ($\rm\ref{XX}$) is Hermitian so there is no exceptional point of the matrix $\textbf{X}$ in the AFM phase. We can easily find that the eigenvalues of the matrix $\textbf{X}$ are two $\beta^{AFM}_{+}$'s and two $\beta^{AFM}_{-}$'s, where
\begin{eqnarray}
\label{betaAM}
\beta^{AFM}_{\pm}=\frac{1}{4}\left(\Gamma_g+\Gamma_l\pm\sqrt{(\Gamma_g-\Gamma_l)^2+4g^2}\right).
\end{eqnarray}
Since $\Gamma_{g}\Gamma_{l}>g^{2}$ for the AM phase, the real part of 
$\beta^{AFM}_{\pm}$ is positive,
\begin{eqnarray}
\textrm{Re}[\beta^{AFM}_{+}]>\textrm{Re}[\beta^{AFM}_{-}]&=&\frac{1}{4}\left(\Gamma_g+\Gamma_l-\sqrt{(\Gamma_g-\Gamma_l)^2+4g^2}\right)\nonumber\\
&>&\frac{1}{4}\left(\Gamma_g+\Gamma_l-\sqrt{(\Gamma_g-\Gamma_l)^2+4\Gamma_g\Gamma_l}\right)\nonumber\\
&=&\frac{1}{4}\left(\Gamma_g+\Gamma_l-(\Gamma_g+\Gamma_l)\right)=0.
\end{eqnarray}
Thus, we can obtain the entire Liouvillian spectrum $\lambda^{AFM}$ as Eq.(\ref{lambdaAM}) by using Theorem 2.
Furthermore, in the limit $g^{2}\to\Gamma_{g}\Gamma_{l}$, we see
\begin{eqnarray}
\label{limit1}
\beta^{AM}_{\pm}\to\frac{1}{4}\left(\Gamma_g+\Gamma_l\pm(\Gamma_g+\Gamma_l)\right)=\frac{1}{2}(\Gamma_g+\Gamma_l),0.
\end{eqnarray}
This shows that the Liouvillian gap is closed at the phase boundary, $\Gamma_{g}\Gamma_{l}=g^{2}$, and many eigenvalues approach 0, which is an anomalous property from the view point of conventional dissipative phase transitions $\text{\cite{Minganti2}}$ as pointed out in Ref.$\text{\cite{Huber1}}$. Also, the time evolution of the normalized magnetizations $\braket{S_{z}}/S$ in the AFM $\ket{\Uparrow\Downarrow}$ can be calculated from Eqs.($\rm\ref{XX}$), ($\rm\ref{AMY}$) and ($\rm\ref{Zt}$).

\subsection{Third quantization for the FM phase}
Next, we apply the third quantization to the FM $\ket{\Uparrow\Uparrow}$ phase. The Hamiltonian and Lindblad equation can be written as 
\begin{eqnarray}
H&=&g(c_A^{\dagger}c_B+\rm{H.c.}),\\
\label{lind2}\dot{\rho}=\hat{\mathcal{L}}\rho&=&-i[H,\rho]+\Gamma_g\mathcal{D}[c_A]\rho+\Gamma_l\mathcal{D}[c_B^{\dagger}]\rho
\end{eqnarray}
with the HP approximation ($\rm\ref{+}$) for $S\gg1$. From the Lindblad equation ($\rm\ref{lind2}$), we can write the dissipation operators ($\rm\ref{dissipation1}$) as
\begin{eqnarray}
L_{1}=\sqrt{\Gamma_{g}}c_{A}=\left(
   \begin{array}{c}
      \sqrt{\Gamma_{g}}  \\
      0
    \end{array}
  \right)\cdot\left(
   \begin{array}{c}
     c_{A}  \\
      c_{B}
    \end{array}
  \right)=\underline{l_{1}}\cdot\underline{c},\\
  L_{2}=\sqrt{\Gamma_{l}}c_{B}^{\dagger}=\left(
   \begin{array}{c}
     0  \\
       \sqrt{\Gamma_{l}}
    \end{array}
  \right)\cdot\left(
   \begin{array}{c}
     c_{A}^{\dagger}  \\
      c_{B}^{\dagger}
    \end{array}
  \right)=\underline{k_{2}}\cdot\underline{c^{\dagger}}.
\end{eqnarray}
Therefore, we can calculate the matrices $\textbf{H}$ in Eq.($\rm\ref{Hamiltonian1}$) and $\textbf{M}$ ($\rm\ref{M1}$), $\textbf{N}$ ($\rm\ref{N1}$) as
\begin{eqnarray}
\textbf{H}=\left(
   \begin{array}{rr}
      0 & g  \\
      g & 0 
    \end{array}
  \right),\ \ \textbf{M}=\left(
   \begin{array}{rr}
      \Gamma_{g}& 0  \\
      0 & 0
    \end{array}
  \right),\ \ \textbf{N}=\left(
   \begin{array}{rr}
      0& 0  \\
      0 & \Gamma_{l}
    \end{array}
  \right)
\end{eqnarray}
and $\textbf{K}=\textbf{L}=\textbf{0}$. Therefore, we find that the matrices $\textbf{X}$ ($\rm\ref{X1}$) and $\textbf{Y}$ ($\rm\ref{Y}$) are written as
\begin{eqnarray}
\label{FMX}
\textbf{X}=\frac{1}{2}\left(
    \begin{array}{rrrr}
     \Gamma_g & ig&0&0\\
      ig& -\Gamma_l &0&0\\
   0 & 0&\Gamma_g&-ig\\
      0&0&-ig& -\Gamma_l 
    \end{array}
  \right)
\end{eqnarray}
and
\begin{eqnarray}
\label{FMY}
\textbf{Y}=\left(
    \begin{array}{rrrr}
     0& 0&0&0\\
     0& 0 &0&\Gamma_l\\
   0 & 0&0&0\\
      0 &\Gamma_l&0& 0 
    \end{array}
  \right),
\end{eqnarray}
respectively. We first discuss the case with $\Gamma_{g}+\Gamma_{l}\neq 2g$, in which
we can easily find that the eigenvalues of the matrix $\textbf{X}$ ($\rm\ref{FMX}$) are two $\beta^{FM\ket{\Uparrow\Uparrow}}_{+}$'s and two $\beta^{FM\ket{\Uparrow\Uparrow}}_{-}$'s, where
\begin{eqnarray}
\label{betaFM1}
\beta^{FM\ket{\Uparrow\Uparrow}}_{\pm}=\frac{1}{4}\left(\Gamma_g-\Gamma_l\pm\sqrt{(\Gamma_g+\Gamma_l)^2-4g^2}\right).
\end{eqnarray}
The eigenvalue structure of the FM $\ket{\Downarrow\Downarrow}$ phase can be obtained by transforming Eq.($\rm\ref{betaFM1}$) as $\Gamma_{g,l}\to-\Gamma_{g,l}$, so it can be found that the eigenvalue structure of the whole FM phase can be written as %The eigenvalue structure of the FM $\ket{\Downarrow\Downarrow}$ phase is equivalent to the one of the FM $\ket{\Uparrow\Uparrow}$ phase which $\Gamma_{g}$ and $\Gamma_{l}$ are interchanged, so it can be found that the eigenvalue structure of the whole FM phase can be written as
\begin{eqnarray}
\label{betaFM}
\beta^{FM}_{\pm}=\frac{1}{4}\left(|\Gamma_g-\Gamma_l|\pm\sqrt{(\Gamma_g+\Gamma_l)^2-4g^2}\right).
\end{eqnarray}
Since $\Gamma_{g}\Gamma_{l}<g^{2}$ in the FM phases, the real part of 
$\beta^{FM}_{\pm}$ is positive. In fact, if $\Gamma_{g}+\Gamma_{l}<2g$, $\textrm{Re}[\beta^{FM}_{\pm}]=|\Gamma_g-\Gamma_l|/4$ while if $\Gamma_{g}+\Gamma_{l}>2g$, we see that
\begin{eqnarray}
\textrm{Re}[\beta^{FM}_{+}]>\textrm{Re}[\beta^{FM}_{-}]&=&\frac{1}{4}\left(|\Gamma_g-\Gamma_l|-\sqrt{(\Gamma_g+\Gamma_l)^2-4g^2}\right)\nonumber\\
&>&\frac{1}{4}\left(|\Gamma_g-\Gamma_l|-\sqrt{(\Gamma_g+\Gamma_l)^2-4\Gamma_g\Gamma_l}\right)\nonumber\\
&=&\frac{1}{4}\left(|\Gamma_g-\Gamma_l|-|\Gamma_g-\Gamma_l|\right)=0.
\end{eqnarray}
Thus, we can obtain the entire Liouvillian spectrum $\lambda^{FM}$ as Eq.($\ref{lambdaFM}$) by using Theorem 2 and Theorem A1.
Furthermore, in the limit $\Gamma_{g}\Gamma_{l}\to g^{2}$, we see that
\begin{eqnarray}
\label{betalimit}
\beta^{FM}_{\pm}\to\frac{1}{4}\left(|\Gamma_g-\Gamma_l|\pm|\Gamma_g-\Gamma_l|\right)=\frac{1}{2}|\Gamma_g-\Gamma_l|,0.
\end{eqnarray}
This shows that the Liouvillian gap is closed at the phase boundary $\Gamma_{g}\Gamma_{l}=g^{2}$ and many eigenvalues approach to 0.
Furthermore, in the limit $|\Gamma_{g}-\Gamma_{l}|\to 0$, we see
\begin{eqnarray}
\beta^{FM}_{\pm}\to\pm\frac{i}{2}\sqrt{g^{2}-\Gamma^{2}},
\end{eqnarray}
where $\Gamma_{g}=\Gamma_{l}=\Gamma$.
This shows that Liouvillian gap is closed at the phase boundary and there exist some pure imaginary eigenvalues in the $\mathcal{PT}$ phase. Furthermore, we set
\begin{eqnarray}
(m_{1}+m_{2})-(m_{3}+m_{4})&=&q,\\
(m_{1}+m_{2}+m_{3}+m_{4})|\Gamma_{g}-\Gamma_{l}|&=&n,
\end{eqnarray}
and consider the limits $m_{i}\to\infty$,\ $|\Gamma_{g}-\Gamma_{l}|\to0$. Then, it can be found that $q\in\mathbb{Z}$ and $n\in\mathbb{R}_{+}$ because $m_{i}$ is an integer and $|\Gamma_{g}-\Gamma_{l}|$ is continuous. After all, eigenvalue structures are equivalent from both sides of the FM phases in the limit to the $\mathcal{PT}$ phase. Also, the time evolution of the normalized magnetizations $\braket{S_{z}}/S$ in the FM $\ket{\Uparrow\Uparrow}$ phase can be calculated from Eqs.($\rm\ref{FMX}$), ($\rm\ref{FMY}$) and ($\rm\ref{Zt}$). 

When the parameters satisfy $\Gamma_{g}+\Gamma_{l}=2g$, the matrix $\textbf{X}$ ($\rm\ref{FMX}$) is not diagonalizable. This line corresponds to exceptional points of the matrix $\textbf{X}$ 
(of order two) and hence also of the Liouvillian, by the arguments given below Theorem A1 in Appendix A. These facts are also mentioned in the main text.   

%A.1
\section{Derivation of Eqs.($\ref{HH}$), ($\ref{NL}$)}
In this appendix, we derive Eqs.($\ref{HH}$), ($\ref{NL}$).
From the criterion of Huber \textit{et al}. of Liouvillian $\mathcal{PT}$ symmetry ($\ref{HuberPT}$), 
we can find that if $L_{\mu}$ is the dissipation operator, $\mathbb{PT}^{\prime}$($L_{\mu}$) is also one of the dissipation operators. Here, let us set $L_{\mu^{\prime}}=\mathbb{PT}^{\prime}$($L_{\mu}$). Then, we can rewrite the Liouvillian $\mathcal{PT}$ symmetry ($\ref{HuberPT}$) as
\begin{eqnarray}
\label{PTH}
\mathbb{PT}(H)&=H,\\
\label{PTL}\mathbb{PT}^{\prime}(L_{\mu})&=L_{\mu^{\prime}}.
\end{eqnarray}
Since the reflection operator $\mathcal{P}$ reverses the order of the index, $i$ $\to$ $n+1-i$, we can express $\mathcal{P}\underline{a}\mathcal{P}^{-1}$ as $P_{n}\underline{a}$, where $P_{n}$ is defined in Eq.($\ref{P}$). Therefore, we find from the definition of the $\mathbb{PT}$ map ($\ref{HuberPT3}$) 
\begin{eqnarray}
\label{471}
\mathbb{PT}(H)&=&\mathcal{P}\bar{H}\mathcal{P}^{-1}\nonumber\\
&=&(P_{n}\underline{a}^{\dagger})\cdot\bar{\textbf{H}}(P_{n}\underline{a})+(P_{n}\underline{a})\cdot\bar{\textbf{K}}(P_{n}\underline{a})+(P_{n}\underline{a}^{\dagger})\cdot\textbf{K}(P_{n}\underline{a}^{\dagger})\nonumber\\
&=&(P_{n}\underline{a}^{\dagger})^{T}\bar{\textbf{H}}(P_{n}\underline{a})+(P_{n}\underline{a})^T\bar{\textbf{K}}(P_{n}\underline{a})+(P_{n}\underline{a}^{\dagger})^{T}\textbf{K}(P_{n}\underline{a}^{\dagger})\nonumber\\
&=&\underline{a}^{\dagger}\cdot (P_{n}\bar{\textbf{H}}P_{n})\underline{a}+\underline{a}\cdot(P_{n}\bar{\textbf{K}}P_{n})\underline{a}+\underline{a}^{\dagger}\cdot(P_{n}\textbf{K}P_{n})\underline{a}^{\dagger},
\end{eqnarray}
where we use $P_{n}=P_{n}^{T}$. 
From Eqs.($\rm\ref{PTH}$), ($\rm\ref{471}$) and Eq.($\rm\ref{Hamiltonian1}$), we can obtain Eq.($\ref{HH}$). In the same way, we can find from Eq.($\rm\ref{dissipation1}$)
\begin{eqnarray}
\label{D1}
\mathbb{PT}^{\prime}(L_{\mu})&=&\underline{l}_{\mu}\cdot(P_{n}\underline{a}^{\dagger})+\underline{k}_{\mu}\cdot(P_{n}\underline{a})\nonumber\\
&=&(P_{n}\underline{l}_{\mu})\cdot\underline{a}^{\dagger}+(P_{n}\underline{k}_{\mu})\cdot\underline{a},\\
\label{D2}L_{\mu^{\prime}}&=&\underline{l}_{\mu^{\prime}}\cdot\underline{a}+\underline{k}_{\mu^{\prime}}\cdot\underline{a}^{\dagger},
\end{eqnarray}
where we assume that $\underline{l}_{\mu}$ and $\underline{k}_{\mu}$ are real since dissipation rates in Eq.($\ref{lindblad}$) are positive real numbers.
From Eqs.($\rm\ref{PTL}$), ($\rm\ref{D1}$) and Eq.($\rm\ref{D2}$), we see
\begin{eqnarray}
\label{ll}
\underline{l}_{\mu^{\prime}}=P_{n}\underline{k}_{\mu},\ \ \ \ \ \underline{k}_{\mu^{\prime}}=P_{n}\underline{l}_{\mu}.
\end{eqnarray}
Also, we can obtain the following relations from Eqs.($\rm\ref{M1}$), ($\rm\ref{N1}$) and ($\rm\ref{ll}$),
\begin{eqnarray}
\label{N11}
\textbf{N}&=&\sum_{\mu,\mu^{\prime}}(\underline{k}_{\mu}\otimes\underline{k}_{\mu}+\underline{k}_{\mu^{\prime}}\otimes\underline{k}_{\mu^{\prime}})=\sum_{\mu}(\underline{k}_{\mu}\otimes\underline{k}_{\mu}+P_{n}\underline{l}_{\mu}\otimes P_{n}\underline{l}_{\mu})\nonumber\\
&=&\sum_{\mu}(\underline{k}_{\mu}\otimes\underline{k}_{\mu}+P_{n}(\underline{l}_{\mu}\otimes \underline{l}_{\mu})P_{n})=\sum_{\mu}P_{n}(P_{n}(\underline{k}_{\mu}\otimes\underline{k}_{\mu})P_{n}+\underline{l}_{\mu}\otimes \underline{l}_{\mu})P_{n}
\end{eqnarray}
and
\begin{eqnarray}
\label{M11}
\textbf{M}=\sum_{\mu,\mu^{\prime}}(\underline{l}_{\mu}\otimes\underline{l}_{\mu}+\underline{l}_{\mu^{\prime}}\otimes\underline{l}_{\mu^{\prime}})=\sum_{\mu}(P_{n}(\underline{k}_{\mu}\otimes\underline{k}_{\mu})P_{n}+\underline{l}_{\mu}\otimes \underline{l}_{\mu}).
\end{eqnarray}
Comparing Eq.($\rm\ref{N11}$) and Eq.($\rm\ref{M11}$), we can obtain the first relation in Eq.($\ref{NL}$).
In the same way, we can obtain the following relations from Eqs.($\rm\ref{L1}$), ($\rm\ref{ll}$)
\begin{eqnarray}
\label{50}
\textbf{L}=\sum_{\mu,\mu^{\prime}}(\underline{l}_{\mu}\otimes\underline{k}_{\mu}+\underline{l}_{\mu^{\prime}}\otimes\underline{k}_{\mu^{\prime}})=\sum_{\mu}(\underline{l}_{\mu}\otimes\underline{k}_{\mu}+P_{n}\underline{k}_{\mu}\otimes P_{n}\underline{l}_{\mu})=\sum_{\mu}(\underline{l}_{\mu}\otimes\underline{k}_{\mu}+P_{n}(\underline{k}_{\mu}\otimes \underline{l}_{\mu})P_{n}),
\end{eqnarray}
and
\begin{eqnarray}
\label{51}
\textbf{L}^{T}=\sum_{\mu,\mu^{\prime}}(\underline{k}_{\mu}\otimes\underline{l}_{\mu}+\underline{k}_{\mu^{\prime}}\otimes\underline{l}_{\mu^{\prime}})
=\sum_{\mu}(\underline{k}_{\mu}\otimes\underline{l}_{\mu}+P_{n}(\underline{l}_{\mu}\otimes \underline{k}_{\mu})P_{n}).
\end{eqnarray} 
Comparing Eqs.($\rm\ref{50}$) and ($\rm\ref{51}$), we can show the second relation in Eq.($\ref{NL}$).

\section{Derivation of the time evolution of one and two point correlation functions}
We derive the time evolution of one and two point correlation functions ($\ref{psit}$), ($\ref{Zt}$) for quadratic bosonic systems.
First, $\underline{\psi}(t)$ ($\ref{psi2}$) and $\textbf{Z}(t)$ ($\ref{Z3}$) can be rewritten as
\begin{eqnarray}
\label{psit2}
\underline{\psi}(t)=((1|\hat{a}_{i}\ket{\rho(t)},(1|\hat{a}^{\dagger}_{i}\ket{\rho(t)})^{T}=((1|\hat{a}_{(0,i)}\ket{\rho(t)},(1|\hat{a}_{(1,i)}\ket{\rho(t)})^{T}
\end{eqnarray}
and 
\begin{eqnarray}
\label{ZZ}
\textbf{Z}(t)=\left(
   \begin{array}{cc}
    (1|\hat{a}_i\hat{a}_j
\ket{\rho(t)}  &  (1|\hat{a}_{j}^{\dagger}\hat{a}_i
\ket{\rho(t)} \\
  (1|\hat{a}_{i}^{\dagger}\hat{a}_j
\ket{\rho(t)} &  (1|\hat{a}_{i}^{\dagger}\hat{a}_j^{\dagger}
\ket{\rho(t)}
   \end{array}
  \right)=\left(
   \begin{array}{cc}
    (1|\hat{a}_{(0,i)}\hat{a}_{(0,j)}
\ket{\rho(t)}  &  (1|\hat{a}_{(0,i)}\hat{a}_{(1,j)}
\ket{\rho(t)} \\
   (1|\hat{a}_{(1,i)}\hat{a}_{(0,j)}
\ket{\rho(t)}  &  (1|\hat{a}_{(1,i)}\hat{a}_{(1,j)}
\ket{\rho(t)} 
   \end{array}
  \right)
\end{eqnarray}
where $i,j=1,2,..,n$ and we use Eq.($\rm\ref{inner}$). Then, the element of $\underline{\psi}(t)$ and $\textbf{Z}(t)$ can be written as
\begin{eqnarray}
\label{psi}\psi_{(\nu,i)}(t)=(1|\hat{a}_{(\nu,i)}\ket{\rho(t)},\ \ \ \ \ \ \ \ \\
\label{Z2}Z_{(\nu,i),(\mu,j)}(t)=(1|\hat{a}_{(\nu,i)}\hat{a}_{(\mu,j)}
\ket{\rho(t)}
\end{eqnarray}
respectively with $\nu$, $\mu$=0, 1. Note that $\textbf{Z}(t)$ is an extended expression including the time dependence of the matrix $\textbf{Z}$ in Eq.($\rm\ref{Lyapunov}$). The time evolution of the density operator was defined in Eq.($\ref{120}$) as
\begin{eqnarray}
\ket{\rho(t)}=e^{\hat{\mathcal{L}}t}\ket{\rho(0)}.
\end{eqnarray}
Therefore, we can rewrite $\psi_{(\nu,i)}(t)$ in Eq.($\rm\ref{psi}$) and $Z_{(\nu,i),(\mu,j)}(t)$ in Eq.($\rm\ref{Z2}$) as
\begin{eqnarray}
\psi_{(\nu,i)}(t)&=&(1|\hat{a}_{(\nu,i)}\ket{\rho(t)}=(1|\hat{a}_{(\nu,i)}e^{\hat{\mathcal{L}}t}\ket{\rho(0)}\nonumber\\
&=&(1|e^{-\hat{\mathcal{L}}t}\hat{a}_{(\nu,i)}e^{\hat{\mathcal{L}}t}\ket{\rho(0)}=(1|\hat{a}_{(\nu,i)}(t)\ket{\rho(0)},\\
Z_{(\nu,i),(\mu,j)}(t)&=&(1|\hat{a}_{(\nu,i)}\hat{a}_{(\mu,j)}e^{\hat{\mathcal{L}}t}\ket{\rho(0)}=(1|\hat{a}_{(\nu,i)}(t)\hat{a}_{(\mu,j)}(t)\ket{\rho(0)}
\end{eqnarray}
respectively, where we use the stationarity of the left vacuum $(1|e^{-\hat{\mathcal{L}}t}=1$, and we have defined a time evolution of a super-Heisenberg picture by $\hat{a}_{(\nu,j)}(t):=e^{-\hat{\mathcal{L}}t}\hat{a}_{(\nu,j)}e^{\hat{\mathcal{L}}t}$. Next, we calculate the time derivative of $\hat{a}_{(\nu,i)}(t)$ as
\begin{eqnarray}
\label{496}
\frac{d\hat{a}_{(\nu,i)}(t)}{dt}&=&-\hat{\mathcal{L}}e^{-\hat{\mathcal{L}}t}\hat{a}_{(\nu,j)}e^{\hat{\mathcal{L}}t}+e^{-\hat{\mathcal{L}}t}\hat{a}_{(\nu,j)}\hat{\mathcal{L}}e^{\hat{\mathcal{L}}t}=e^{-\hat{\mathcal{L}}t}[\hat{a}_{(\nu,i)},\hat{\mathcal{L}}]e^{\hat{\mathcal{L}}t}\nonumber\\
&=&e^{-\hat{\mathcal{L}}t}\sum_{\mu,\xi=0}^{1}\sum_{j,k=1}^{n}[-(X_{(\mu,j),(\xi,k)}(\hat{a}_{(\nu,i)}\hat{a}_{(\mu,j)}\hat{a}^{\prime}_{(\xi,k)}-\hat{a}_{(\mu,j)}\hat{a}^{\prime}_{(\xi,k)}\hat{a}_{(\nu,i)}))\nonumber\\
&&-(X^{T}_{(\mu,j),(\xi,k)}(\hat{a}_{(\nu,i)}\hat{a}^{\prime}_{(\mu,j)}\hat{a}_{(\xi,k)}-\hat{a}^{\prime}_{(\mu,j)}\hat{a}_{(\xi,k)}\hat{a}_{(\nu,i)}))+(Y_{(\mu,j),(\xi,k)}(\hat{a}_{(\nu,i)}\hat{a}^{\prime}_{(\mu,j)}\hat{a}^{\prime}_{(\xi,k)}-\hat{a}^{\prime}_{(\mu,j)}\hat{a}^{\prime}_{(\xi,k)}\hat{a}_{(\nu,i)}))]e^{\hat{\mathcal{L}}t}.\nonumber\\
\end{eqnarray}
Furthermore, using commutation relations ($\rm\ref{CCR}$), we can calculate Eq.($\rm\ref{496}$) as
\begin{eqnarray}
\frac{d\hat{a}_{(\nu,i)}(t)}{dt}&=&e^{-\hat{\mathcal{L}}t}\sum_{\mu,\xi=0}^{1}\sum_{j,k=1}^{n}[-X_{(\mu,j),(\xi,k)}\hat{a}_{(\mu,j)}\delta_{\nu,\xi}\delta_{i,k}-X^{T}_{(\mu,j),(\xi,k)}\hat{a}_{(\xi,k)}\delta_{\nu,\mu}\delta_{i,j}\nonumber\\
&&+Y_{(\mu,j),(\xi,k)}(\hat{a}^{\prime}_{(\xi,k)}\delta_{\nu,\mu}\delta_{i,j}+\hat{a}^{\prime}_{(\mu,j)}\delta_{\nu,\xi}\delta_{i,k})]e^{\hat{\mathcal{L}}t}\nonumber\\
&=&\sum_{\mu=0}^{1}\sum_{j=1}^{n}[-X_{(\mu,j),(\nu,i)}\hat{a}_{(\mu,j)}(t)-X^{T}_{(\nu,i),(\mu,j)}\hat{a}_{(\mu,j)}(t)+Y_{(\mu,j),(\nu,i)}\hat{a}^{\prime}_{(\mu,j)}(t)+Y_{(\nu,i),(\mu,j)}\hat{a}^{\prime}_{(\mu,j)}(t)].
\end{eqnarray}
Since the matrix $\textbf{Y}$ ($\rm\ref{Y}$) is a symmetric matrix, namely $\textbf{Y}=\textbf{Y}^{T}$, we can rewrite simply as
\begin{eqnarray}
\label{amu}
\frac{d\hat{a}_{(\nu,i)}(t)}{dt}=-2\sum_{\mu=0}^{1}\sum_{j=1}^{n}[X^{T}_{(\nu,i),(\mu,j)}\hat{a}_{(\mu,j)}(t)-Y_{(\nu,i),(\mu,j)}\hat{a}^{\prime}_{(\mu,j)}(t)].
\end{eqnarray}
Therefore, we can obtain the time derivative of the one point correlation function $\psi_{(\nu,i)}(t)$ in Eq.($\rm\ref{psi}$) as
\begin{eqnarray}
\label{E11}
\frac{d\psi_{(\nu,i)}(t)}{dt}&=&\frac{d}{dt}(1|\hat{a}_{(\nu,i)}(t)\ket{\rho(0)}=(1|\frac{d\hat{a}_{(\nu,i)}(t)}{dt}\ket{\rho(0)}\nonumber\\
&=&(1|-2\sum_{\mu=0}^{1}\sum_{j=1}^{n}[X^{T}_{(\nu,i),(\mu,j)}\hat{a}_{(\mu,j)}(t)-Y_{(\nu,i),(\mu,j)}\hat{a}^{\prime}_{(\mu,j)}(t)]\ket{\rho(0)}\nonumber\\
&=&-2\sum_{\mu=0}^{1}\sum_{j=1}^{n}X^{T}_{(\nu,i),(\mu,j)}(1|\hat{a}_{(\mu,j)}(t)\ket{\rho(0)}\nonumber\\
&=&-2\sum_{\mu=0}^{1}\sum_{j=1}^{n}X^{T}_{(\nu,i),(\mu,j)}\psi_{(\mu,j)}(t),
\end{eqnarray}
where we use Eq.($\rm\ref{amu}$) and the relations,
\begin{eqnarray}
(1|\hat{a}^{\prime}_{(0,j)}\ket{\rho(t)}&=&(1|\hat{a^{\dagger}}^{L}_{j}-\hat{a^{\dagger}}^{R}_{j}\ket{\rho(t)}=0,\\
(1|\hat{a}^{\prime}_{(1,j)}\ket{\rho(t)}&=&(1|\hat{a}^{R}_{j}-\hat{a}^{L}_{j}\ket{\rho(t)}=0.
\end{eqnarray}
By integrating Eq.($\rm\ref{E11}$), the time derivative of $\underline{\psi}(t)$ in Eq.($\rm\ref{psit2}$) is found to be give by Eq.($\rm\ref{psit}$).

Next, we calculate the time evolution of two point correlation functions. The time derivative of the two point correlation function $Z_{(\nu,i),(\mu,j)}$ in Eq.($\rm\ref{Z2}$) is written as
\begin{eqnarray}
\label{E14}
\frac{dZ_{(\nu,i),(\mu,j)}(t)}{dt}&=&(1|\frac{d\hat{a}_{(\nu,i)}}{dt}\hat{a}_{(\mu,j)}+\hat{a}_{(\nu,i)}\frac{d\hat{a}_{(\mu,j)}}{dt}\ket{\rho(0)}\nonumber\\
&=&-2(1|\sum_{\xi=0}^{1}\sum_{k=1}^{n}[X^{T}_{(\nu,i),(\xi,k)}\hat{a}_{(\xi,k)}-Y_{(\nu,i),(\xi,k)}\hat{a}^{\prime}_{(\xi,k)}]\hat{a}_{(\mu,j)}
+\hat{a}_{(\nu,i)}[X^{T}_{(\mu,j),(\xi,k)}\hat{a}_{(\xi,k)}-Y_{(\mu,j),(\xi,k)}\hat{a}^{\prime}_{(\xi,k)}]\ket{\rho(0)}\nonumber\\
&=&-2\sum_{\xi=0}^{1}\sum_{k=1}^{n}[X^{T}_{(\nu,i),(\xi,k)}Z_{(\xi,k),(\mu,j)}+X^{T}_{(\mu,j),(\xi,k)}Z_{(\nu,i),(\xi,k)}
-Y_{(\mu,j),(\xi,k)}\delta_{\nu,\xi}\delta_{i,k}]\nonumber\\
&=&-2(\textbf{X}^{T}\textbf{Z}(t)+\textbf{Z}(t)\textbf{X})_{(\nu,i),(\mu,j)}+2Y_{(\nu,i),(\mu,j)},
\end{eqnarray}
where we use the relations,
\begin{eqnarray}
(1|\hat{a}_{(\nu,i)}\hat{a}^{\prime}_{(\mu,j)}\ket{\rho(t)}=\delta_{\nu,\mu}\delta_{i,j},\\
(1|\hat{a}^{\prime}_{(\nu,i)}\hat{a}_{(\mu,j)}\ket{\rho(t)}=0.\ \ \ \ \ \ 
\end{eqnarray}
By integrating Eq.($\rm\ref{E14}$), the time derivative of $\textbf{Z}(t)$ in Eq.($\rm\ref{ZZ}$) is found to be given by Eq.($\ref{Zt}$).

\subsection{The open 2-boson model with balanced gain and loss}
In this section, we provide the expression of the time dependence of one and two point correlation functions for the open 2-boson model in section IV.C. 
From Eq.($\ref{bosonn}$), we can find that the matrices $\textbf{X}$ ($\rm\ref{X1}$) and $\textbf{Y}$ ($\rm\ref{Y}$) are written as
\begin{eqnarray}
\label{bosonX}
\textbf{X}=\frac{1}{2}\left(
    \begin{array}{rrrr}
     \Gamma & ig&0&0\\
      ig& -\Gamma &0&0\\
   0 & 0&\Gamma&-ig\\
      0&0&-ig& -\Gamma 
    \end{array}
  \right)
\end{eqnarray}
and
\begin{eqnarray}
\label{bosonY}
\textbf{Y}=\left(
    \begin{array}{rrrr}
     0& 0&0&0\\
     0& 0 &0&\Gamma\\
   0 & 0&0&0\\
      0 &\Gamma&0& 0 
    \end{array}
  \right),
\end{eqnarray}
respectively. From Eqs.($\rm\ref{bosonX}$), ($\ref{psit}$), we can derive the time dependence of the one point correlation functions as
\begin{eqnarray}
\braket{c_{A}}=a_{1}e^{i\sqrt{g^{2}-\Gamma^{2}}t}+a_{2}e^{-i\sqrt{g^{2}-\Gamma^{2}}t},\\
\braket{c_{B}}=a_{3}e^{i\sqrt{g^{2}-\Gamma^{2}}t}+a_{4}e^{-i\sqrt{g^{2}-\Gamma^{2}}t},
\end{eqnarray}
where $a_{1}, a_{2}$ are constant and
\begin{eqnarray}
a_{3}=\frac{i\sqrt{g^{2}-\Gamma^{2}}+\Gamma}{ig}a_{1},\ \ \ \ a_{4}=\frac{-i\sqrt{g^{2}-\Gamma^{2}}+\Gamma}{ig}a_{2}.
\end{eqnarray}
From Eqs.($\rm\ref{bosonX}$), ($\rm\ref{bosonY}$) and ($\rm\ref{Zt}$), we can also derive the time dependence of the two point correlation functions as
\begin{eqnarray}
\braket{c_{A}^{\dagger}c_{A}}=b_{1}e^{2i\sqrt{g^{2}-\Gamma^{2}}t}+b_{2}e^{-2i\sqrt{g^{2}-\Gamma^{2}}t}+b_{3}t+b_{4},\\
\braket{c_{B}^{\dagger}c_{B}}=b_{5}e^{2i\sqrt{g^{2}-\Gamma^{2}}t}+b_{6}e^{-2i\sqrt{g^{2}-\Gamma^{2}}t}+b_{7}t+b_{8},
\end{eqnarray}
where $b_{1}, b_{2}, b_{3}, b_{4}$ are constant and 
\begin{eqnarray}
b_{5}&=&\frac{i\sqrt{g^{2}-\Gamma^{2}}+\Gamma}{i\sqrt{g^{2}-\Gamma^{2}}-\Gamma}b_{1},\ \ \ \ 
b_{6}=\frac{-i\sqrt{g^{2}-\Gamma^{2}}+\Gamma}{i\sqrt{g^{2}-\Gamma^{2}}+\Gamma}b_{2}, \\
b_{7}&=&b_{3},\ \ \ \ 
b_{8}=b_{4}-2+b_{3}/\Gamma.
\end{eqnarray}
 We can easily find that the dynamics clearly changes at $\Gamma=g$.
 
 %$S^z_A(t)$
 
 %\subsection{The open 2-mode Rabi model with balanced gain and loss}
%The matrix $\textbf{X}$ ($\ref{X}$) can be derived as 
%\begin{eqnarray}
%\label{X11}
%\textbf{X}=\frac{1}{2}\left(
  % \begin{array}{cccc}
  %   i\omega-ir+\Gamma & -ir&ir&ir  \\
 %    -ir& i\omega-ir-\Gamma &ir&ir  \\
%    -ir&-ir& -i\omega+ir+\Gamma &ir  \\
 %   -ir& -ir&ir& -i\omega+ir-\Gamma  
%   \end{array}
%  \right),
%\end{eqnarray}
%where $r=\omega g^{2}/2$.

%\begin{thebibliography}{99}
%\bibitem{Ashidasan}Ashida Y., Gong, Z. $\&$ Ueda, M. Non-Hermitian Physics. arXiv 1-191 (2020).
%\end{thebibliography}


\begin{thebibliography}{99}

\bibitem{Huberr}S. D. Huber, Topological mechanics, Nat. Phys. $\textbf{12}$, 621 (2016).

\bibitem{Ruschhaupt}A. Ruschhaupt, F. Delgado,  $\&$ J. G. Muga, Physical realization of $\mathcal{PT}$-symmetric potential scattering in a planar slab waveguide, J. Phys. A. Math. Gen. $\textbf{38}$, L171 (2005).

\bibitem{Joglekar}Y. N. Joglekar, $\&$ S. J. Wolf, The elusive memristor: Properties of basic electrical circuits, Eur. J. Phys. $\textbf{30}$, 661 (2009).



\bibitem{Ashidasan}Y. Ashida, Z. Gong, and M. Ueda, Non-Hermitian physics, Adv. Phys. $\textbf{69}$, 3 (2020).

\bibitem{BenderC.M.Boettcher} C. M. Bender, $\&$ S. Boettcher,  Real spectra in non-hermitian hamiltonians having $\mathcal{PT}$ symmetry. Phys. Rev. Lett. $\textbf{80}$, 5243 (1998).

\bibitem{Bender}C. M. Bender, B. K. Berntson, D. Parker, $\&$ E. Samuel, Observation of $\mathcal{PT}$ phase transition in a simple mechanical system, Am. J. Phys. $\textbf{81}$, 173 (2013).

\bibitem{RoterC}C. E. R$\ddot{\textrm{u}}$ter, K. G. Makris, R. El-Ganainy, D. N. Christodoulides, M. Segev, and D.
Kip, Observation of parity-time symmetry in optics, Nat. Phys. $\textbf{6}$, 192 (2010).

%\bibitem{Xu}Xu, H., Mason, D., Jiang, L. $\&$ Harris, J. G. E. Topological energy transfer in an optomechanical system with exceptional points. Nature $\textbf{537}$, 80-83 (2016).

\bibitem{Alaeian}H. Alaeian,  $\&$ J. A. Dionne, Parity-time-symmetric plasmonic
metamaterials, Phys. Rev. A $\textbf{89}$, 033829 (2014).

\bibitem{Schindler}J. Schindler, A. Li, M. C. Zheng, F. M. Ellis, $\&$ T. Kottos, Experimental study of active $LRC$ circuits with $\mathcal{PT}$ symmetries, Phys. Rev. A $\textbf{84}$, 040101(R) (2011).


%\bibitem{Regensburger}Regensburger, A. et al. Parity-time synthetic photonic lattices. Nature $\textbf{488}$, 167-171 (2012).

\bibitem{Makris}K. G. Makris, R. El-Ganainy, D. N. Christodoulides,  $\&$ Z. H. Musslimani, Beam dynamics in $\mathcal{PT}$ symmetric optical lattices, Phys. Rev. Lett. $\textbf{100}$, 103904 (2008).

\bibitem{Chen}W. Chen, J. Zhang, B. Peng, S. K. Ozdemir, X. Fan, and L. Yang, Parity-time-symmetric whispering-gallery mode nanoparticle sensor, Photonics Res. $\textbf{6}$, A23 (2018).

%\bibitem{Chen}Chen, W., $\ddot{\textrm{O}}$zdemir, S. K., Zhao, G., Wiersig, J. $\&$ Yang, L. Exceptional points enhance sensing in an optical microcavity. Nature $\textbf{548}$, 192-195 (2017).

\bibitem{Guo}A. Guo, J. Salamo, D. Duchesne, R. Morandotti, M. Volatier-Ravat, V. Aimez, G. A. Siviloglou, $\&$ D. N. Christodoulides, Observation of $\mathcal{PT}$-symmetry breaking in complex optical potentials, Phys. Rev. Lett. $\textbf{103}$, 093902 (2009).


%\bibitem{Feng}Feng, L., Wong, Z. J., Ma, R. M., Wang, Y. $\&$ Zhang, X. Single-mode laser by parity-time symmetry breaking. Science. $\textbf{346}$, 972-975 (2014).

\bibitem{Ramezani}H. Ramezani, T. Kottos, R. El-Ganainy, $\&$ D. N. Christodoulides,  Unidirectional nonlinear $\mathcal{PT}$-symmetric optical structures, Phys. Rev. A $\textbf{82}$, 043803 (2010).




% V. Gorini, A. Kossakowski, and E. C. G. Sudarshan, J. Math. Phys. $\textbf{17}$, 821 (1976).

\bibitem{Lindbladref}G. Lindblad, Comm. Math. Phys. $\textbf{48}$, 119 (1976).



\bibitem{Breuer}H. Breuer and F. Petruccione, \textit{The Theory of Open Quantum Systems} (Oxford University
Press, Oxford, 2007).


\bibitem{ARivas}A. Rivas and S. F. Huelga, \textit{Open Quantum Systems: An Introduction}, SpringerBriefs in Physics (Springer, Heidelberg, 2012).

\bibitem{Minganti1}F. Minganti, A. Miranowicz, R. W. Chhajlany, $\&$ F. Nori, Quantum exceptional points of non-Hermitian Hamiltonians and Liouvillians: The effects of quantum jumps, Phys. Rev. A $\textbf{100}$, 062131 (2019).

\bibitem{MingantiH}F. Minganti, A. Miranowicz, R. W. Chhajlany, I. I. Arkhipov, $\&$ Franco Nori, Hybrid-Liouvillian formalism connecting exceptional points of non-Hermitian Hamiltonians and Liouvillians via postselection of quantum trajectories, Phys. Rev. A $\textbf{101}$, 062112 (2020).



\bibitem{Wu1}Y. Wu, W. Liu, J. Geng, X. Song, X. Ye, C.-K. Duan, X. Rong, and J. Du, Observation of parity-time symmetry breaking in a single-spin system, Science $\textbf{364}$, 878 (2019).

\bibitem{Naghiloo}M. Naghiloo, M. Abbasi, Y. N. Joglekar, $\&$ K. W. Murch,  Quantum state tomography across the exceptional point in a single dissipative qubit, Nat. Phys. $\textbf{15}$, 1232 (2019).

%dissipative phase transition

\bibitem{Minganti2}F. Minganti, A. Biella, N. Bartolo, $\&$ C. Ciuti, Spectral theory of Liouvillians for dissipative phase transitions, Phys. Rev. A $\textbf{98}$, 042118 (2018).


\bibitem{Kessler}E. M. Kessler, G. Giedke, A. Imamoglu, S. F. Yelin, M. D. Lukin, $\&$ J. I. Cirac, Dissipative phase transition in a central spin system, Phys. Rev. A $\textbf{86}$, 012116 (2012).




\bibitem{Hwang}M. J. Hwang, P. Rabl, $\&$ M. B. Plenio, Dissipative phase transition in the open quantum Rabi model, Phys. Rev. A $\textbf{97}$, 013825 (2018).


\bibitem{Casteels1}W. Casteels, R. Fazio, $\&$ C. Ciuti, Critical dynamical properties of a first-order dissipative phase transition, Phys. Rev. A $\textbf{95}$, 012128 (2017).


\bibitem{Rota1}R. Rota, F. Storme, N. Bartolo, R. Fazio, $\&$ C. Ciuti, Critical behavior of dissipative two-dimensional spin lattices, Phys. Rev. B $\textbf{95}$, 134431 (2017).



\bibitem{Lee1}T. E. Lee, S. Gopalakrishnan, $\&$ M. D. Lukin, Unconventional magnetism via optical pumping of interacting spin systems, Phys. Rev. Lett. $\textbf{110}$, 257204 (2013).



\bibitem{Scheel}S. Scheel, $\&$ A. Szameit, $\mathcal{PT}$-symmetric photonic quantum systems with gain and loss do not exist, Eur. Phys. Lett. $\textbf{122}$, 34001 (2018).
%Europhys. Lett. 122, 34001 (2018).

\bibitem{Prosen1}T. Prosen,  $\mathbb{PT}$-symmetric quantum Liouvillian dynamics, Phys. Rev. Lett. $\textbf{109}$, 090404 (2012).

\bibitem{Huber2}J. Huber, P. Kirton, S. Rotter,  $\&$ P. Rabl, Emergence of $\mathcal{PT}$-symmetry breaking in open quantum systems, SciPost Phys. $\textbf{9}$, 52 (2020).

%SciPost Phys. 9, 52 (2020).

\bibitem{Huber1}J. Huber, P. Kirton, $\&$ P. Rabl,  Nonequilibrium magnetic phases in spin lattices with gain and loss, Phys. Rev. A $\textbf{102}$, 012219 (2020).

\bibitem{Kepesidis}K. V. Kepesidis, T. J. Milburn, J. Huber, K. G. Makris, S. Rotter, and P. Rabl, $\mathcal{PT}$-symmetry breaking in the steady state of microscopic gain-loss systems, New J. Phys. $\textbf{18}$, 095003 (2016).

\bibitem{Prosen3}T. Prosen, Third quantization: A general method to solve master equations for quadratic open Fermi systems, New J. Phys. $\textbf{10}$, 043026 (2008).

\bibitem{Prosen4}T. Prosen, $\&$ T. H. Seligman, Quantization over boson operator spaces, J. Phys. A Math. Theor. $\textbf{43}$, 392004 (2010).

\bibitem{Holstein}T. Holstein, and H. Primakoff, Field dependence of the intrinsic domain magnetization of a ferromagnet, Phys. Rev. $\textbf{58}$,
1098 (1940).


%\bibitem{Albert}Albert, V. V. $\&$ Jiang, L. Symmetries and conserved quantities in Lindblad master equations. Phys. Rev. A - At. Mol. Opt. Phys. $\textbf{89}$, (2014).





\bibitem{MostafazadehA1}A. Mostafazadeh,  Pseudo-Hermiticity versus PT symmetry: The necessary condition for the reality of the spectrum of a non-Hermitian Hamiltonian, J. Math. Phys. $\textbf{43}$, 205 (2002).




\bibitem{Prosen2}T. Prosen, Generic examples of $\mathbb{PT}$-symmetric qubit (spin-1/2) Liouvillian dynamics, Phys. Rev. A $\textbf{86}$, 044103 (2012).

\bibitem{Huybrechts}D. Huybrechts, F. Minganti, F. Nori, M. Wouters, $\&$ N. Shammah, Validity of mean-field theory in a dissipative critical system: Liouvillian gap, $\mathbb{PT}$-symmetric antigap, and permutational symmetry in the $XYZ$ model, Phys. Rev. B $\textbf{101}$, 214302 (2020).

\bibitem{Van}M. van Caspel, $\&$ V. Gritsev, Symmetry-protected coherent relaxation of open quantum systems, Phys. Rev. A $\textbf{97}$, 052106 (2018).


\bibitem{Nigro}Nigro, D. On the uniqueness of the steady-state solution of the Lindblad-Gorini-Kossakowski-Sudarshan equation, J. Stat. Mech. (2019), 043202.


\bibitem{Heiss}W. D. Heiss, The physics of exceptional points, J. Phys. A Math. Theor. $\textbf{45}$ 444016 (2012).

\bibitem{Kanki}K. Kanki, S. Garmon, S. Tanaka, $\&$ T. Petrosky, Exact description of coalescing eigenstates in open quantum systems in terms of microscopic Hamiltonian dynamics, J. Math. Phys. $\textbf{58}$, 092101 (2017).



\bibitem{Daley}A. J. Daley, Quantum trajectories and open many-body quantum systems, Adv. Phys. $\textbf{63}$, 77 (2014).

\bibitem{Johansson1}J. Johansson, P. Nation, and F. Nori, QuTiP: An open-source Python framework for the dynamics of open quantum systems, Comp. Phys. Commun. $\textbf{183}$, 1760 (2012).

\bibitem{Johansson2}J. Johansson, P. Nation, and F. Nori, QuTiP 2: A Python framework for the dynamics of open quantum systems, Comp. Phys. Commun. $\textbf{184}$, 1234 (2013).

\bibitem{Roccati1}F. Roccati, S. Lorenzo, G. M. Palma, G. T. Landi, M. Brunelli, and F. Ciccarello, Quantum correlations in $\mathcal{PT}$-symmetric systems, Quantum Sci. Technol. $\textbf{6}$ 025005 (2021).

\bibitem{Purkayastha}A. Purkayastha, M. Kulkarni, and Y. N. Joglekar,  Emergent  $\mathcal{PT}$ Symmetry in a double-quantum-dot circuit QED setup, Phys. Rev. Res. $\textbf{2}$, 043075 (2020).



\bibitem{Arkhipov1}I. I. Arkhipov, A. Miranowicz, F. Minganti, $\&$ F. Nori, Quantum and semiclassical exceptional points of a linear system of coupled cavities with losses and gain within the Scully-Lamb laser theory, Phys. Rev. A $\textbf{101}$, 013812 (2020).


\bibitem{Davis}J. M. Davis,  I. A. Gravagne, R. J. Marks, $\&$ A. A. Ramos, \textit{Proceedings of the 42nd Meeting of the Southeastern Symposium on System Theory, Tyler, 2010}
(IEEE, Piscataway, 2010), p. 329.

%72
\bibitem{Behr}M. Behr, P. Benner, $\&$ J. Heiland, Solution formulas for differential Sylvester and Lyapunov equations, Calcolo $\textbf{56}$, 51 (2019).



%M. Behr, P. Benner, and J. Heiland. \Solution formulas for differential Sylvester and Lyapunov equations". J. Calcolo 56 (2019).

%\bibitem{MostafazadehA1}Mostafazadeh, A. Pseudo-Hermiticity versus PT symmetry: The necessary condition for the reality of the spectrum of a non-Hermitian Hamiltonian. J. Math. Phys. $\textbf{43}$, 205-214 (2002).




%\bibitem{Li1}Li, J. et al. Observation of parity-time symmetry breaking transitions in a dissipative Floquet system of ultracold atoms. Nat. Commun. $\textbf{10}$, 1-7 (2019).

%\bibitem{Brody}Brody, D. C. $\&$ Graefe, E. M. Mixed-state evolution in the presence of gain and loss. Phys. Rev. Lett. 109, 1-5 (2012).



%\bibitem{Prosen2}Prosen, T. Generic examples of PT-symmetric qubit (spin-1/2) Liouvillian dynamics. Phys. Rev. A At. Mol. Opt. Phys. $\textbf{86}$, 1-2 (2012).

%\bibitem{Iemin}Iemini, F. et al. Boundary Time Crystals. Phys. Rev. Lett. $\textbf{121}$, 35301 (2018).

%\bibitem{Tucker}Tucker, K. et al. Shattered time: Can a dissipative time crystal survive many-body correlations? New J. Phys. (2018).

%\bibitem{MingantiA}, F., Arkhipov, I. I., Miranowicz, A. $\&$ Nori, F. Correspondence between dissipative phase transitions of light and time crystals. arXiv 1-10 (2020).




\bibitem{Arkhipov}I. I. Arkhipov, A. Miranowicz, F. Minganti, $\&$ F. Nori, Liouvillian exceptional points of any order in dissipative linear bosonic systems: Coherence functions and switching between $\mathcal{PT}$ and anti-$\mathcal{PT}$ symmetries, Phys. Rev. A $\textbf{102}$, 033715 (2020).

\bibitem{Arkhipov3}I. I. Arkhipov, F. Minganti, A. Miranowicz, $\&$ F. Nori, Generating high-order quantum exceptional points, Phys. Rev. A $\textbf{104}$, 012205


\bibitem{a}In a way similar to Theorem 1, the matrix $i\textbf{X}$ has conventional anti-$\mathcal{PT}$ symmetry, $\{i\textbf{X},PT\}=(i\textbf{X})PT+PT(i\textbf{X})=0$, if the model satisfies 
$\hat{\mathcal{L}}[\mathbb{PT}(H);\mathbb{PT}^\prime(L_\mu),\mu=1,2,\cdots]=\hat{\mathcal{L}}[-H;L_\mu^{\dagger},\mu=1,2,\cdots]$.




%\bibitem{QED} F. Quijandr$\acute{\text{i}}$a, U. Naether, Sahin K. $\ddot{\text{O}}$zdemir, F. Nori, and D. Zueco $\mathcal{PT}$-symmetric circuit QED, Phys. Rev. A $\textbf{97}$, 053846 (2018).


\bibitem{Dast1} D. Dast, D. Haag, H. Cartarius, $\&$ G. Wunner, Quantum master equation with balanced gain and loss. Phys. Rev. A $\textbf{90}$, 052120 (2014).

\bibitem{Braak}D. Braak, Integrability of the Rabi Model,
Phys. Rev. Lett. $\textbf{107}$, 100401 (2011).

\bibitem{Hwang2}M. J. Hwang, R. Puebla, and M. B. Plenio, Quantum Phase Transition and Universal Dynamics in the Rabi Model, Phys.
Rev. Lett. $\textbf{115}$, 180404 (2015).

\bibitem{Puebla}R. Puebla, M. J. Hwang, and M. B. Plenio, Excited-state quantum phase transition in the Rabi model, Phys. Rev. A $\textbf{94}$, 023835 (2016).


\bibitem{Zhang} Y. Z. Zhang, On the solvability of the quantum Rabi model and its 2-photon and two-mode generalizations,
J. Math. Phys. $\textbf{54}$, 102104 (2013).

\bibitem{Cai}M.-L. Cai, Z.-D. Liu, W.-D. Zhao, Y.-K. Wu, Q.-X. Mei,
Y. Jiang, L. He, X. Zhang, Z.-C. Zhou, and L.-M. Duan, Observation of a quantum phase transition in the quantum Rabi model with a single trapped ion, Nat. Commun. $\textbf{12}$, 1126 (2021).



%\bibitem{Hwang1}M. J. Hwang, P. Rabl, M. B. Plenio, Dissipative phase transition inthe open quantum Rabi model, Phys Rev A. $\textbf{97}$ 013825 (2018).



\bibitem{Malekakhlagh}M. Malekakhlagh and A. W. Rodriguez, Quantum Rabi Model with Two-Photon Relaxation,
Phys. Rev. Lett. $\textbf{122}$, 043601

\bibitem{Wang}N. Wang, Z. R. Gong, J. Lu, and L. Zhou, Phases Transitions in a cross-cavity quantum Rabi model possessing PT symmetric structure, Front. Phys. $\textbf{7}$, 127 (2019).


\bibitem{Jan} J. Wiersig, Robustness of exceptional-point-based sensors against parametric noise: The role of Hamiltonian and Liouvillian degeneracies, Phys. Rev. A $\textbf{101}$, 053846 (2020).

\bibitem{He}Y. He and C. C. Chien, Comparison of topological classifications of quadratic bosonic excitations with examples, arXiv:2103.15200.

\bibitem{van}M. van Caspel, S. E. T. Arze, $\&$ I. P. Castillo, Dynamical signatures of topological order in the driven-dissipative
Kitaev chain, SciPost Phys. $\textbf{6}$, 26 (2019).

\bibitem{Dangel}F. Dangel, M. Wagner,  H. Cartarius, J. Main, $\&$ G. Wunner, Topological invariants in dissipative extensions of the Su-Schrieffer-Heeger model, Phys. Rev. A $\textbf{98}$, 013628 (2018).


\bibitem{Lieu1}S. Lieu, M. McGinley, $\&$ N. R. Cooper, Tenfold Way for Quadratic Lindbladians, Phys. Rev. Lett. $\textbf{124}$, 040401 (2020).

%Braak,Hwang,Puebla,Zhang,Cai,Hwang1,Wang




\bibitem{Prosen11}T. Prosen, Spectral theorem for the Lindblad equation for quadratic open fermionic systems, J. Stat. Mech. (2010) P07020.



%\bibitem{yuma}Please see Appendix E.

\end{thebibliography}
\end{document}